\documentclass[aps,twocolumn]{revtex4-1}

\usepackage{graphicx}
\usepackage{amsmath,mathtools,dashbox,amssymb}
\usepackage{xcolor,ulem,esint}

\newcommand*{\descbox}[1]{\mathop{\boxed{#1}}\limits}
\newcommand*{\dashboxed}[1]{\dbox{\ensuremath{#1}}}

\begin{document}
\title{Cotunneling effects in the  geometric statistics of a nonequilibrium  spintronic junction}


\author{Mriganka Sandilya$^a$}
\author{Javed Akhtar$^b$}
\author{Manash Jyoti Sarmah$^b$}
\author{Himangshu Prabal Goswami$^b$}
\email{hpg@gauhati.ac.in}
\affiliation{$^a$Department of Physics, $^b$Department of Chemistry,  Gauhati University, Gopinath Bardoloi Nagar,
Jalukbari, Guwahati-781014, Assam, India}
\date{\today}

\begin{abstract} 

In the nonequilibrium steadystate of electronic transport across a spin-resolved quantronic junction, we investigate the role of cotunneling on the emergent statistics under phase-different adiabatic modulation of the reservoirs' chemical potentials. By explicitly identifying the sequential and inelastic cotunneling rates, we numerically evaluate the geometric or Pancharatnam-Berry contributions to the spin exchange flux. We identify the relevant conditions wherein the sequential and cotunneling processes compete and selectively influence the total geometric flux upshot. The Fock space coherences are found to suppress the cotunneling effects when the system reservoir couplings are comparable. The cotunneling contribution to the total geometric flux can be made comparable to the sequential contribution by creating a rightsided asymmetry in the system-reservoir coupling strength. Using a recently proposed geometric thermodynamic uncertainty relationship, we numerically estimate the total rate of minimal entropy production. The geometric flux and the minimum entropy are found to be nonlinear as a function of the interaction energy of the junction's spin orbitals. 

\end{abstract}

\maketitle





\section{Introduction}
Phase different binary modulation of independent system variables introduces a geometric phase during the adiabatic evolution of isolated and open quantum systems \cite{pancharatnam1956generalized,berry1984quantal,mukunda1993quantum,carollo2003geometric,sinitsyn2007berry,PhysRevLett.104.170601,goswami2016geometric,giri2019nonequilibrium,yuge2012geometrical,wang2022geometric,hino2021geometrical,PhysRevE.96.052129,giri2019nonequilibrium,lu2022geometric}. In the case of non-equilibrium systems, where fermion or bosonic transport happens due to thermal or potential gradient, the geometric phase is exactly not a phase, but a phaselike quantity with a conventional geometric interpretation. It can be simply referred to as its geometricity \cite{PhysRevLett.104.170601,goswami2016geometric,yuge2012geometrical,wang2022geometric,hino2021geometrical,PhysRevE.96.052129,lu2022geometric}. The geometricity is usually quantified by evaluating the cumulant generating function with an additive geometric correction term apart from the dynamic component \cite{sinitsyn2007berry,PhysRevLett.104.170601,goswami2016geometric,PhysRevLett.99.220408}. Such geometric contribution to the quantum statistics introduces  non-triviality in the simplest of non-equilibrium models such as single resonant levels \cite{PhysRevLett.104.170601,goswami2016geometric}. The holonomy in the parameter space changes when the either the temperatures, energies, or chemical potentials in the system or the reservoirs adiabatically evolve in time either in a cyclic or non-cyclic fashion\cite{PhysRevResearch.5.033014}. This allows an additional contribution to the flux and higher-order quanta fluctuations apart from the conventional dynamic component of  non-driven junctions  \cite{PhysRevLett.104.170601,goswami2016geometric,yuge2012geometrical,wang2022geometric,hino2021geometrical,PhysRevE.96.052129,lu2022geometric}. Such adiabatic pumping\cite{PhysRevB.99.035437,PhysRevA.102.043512} was found to be fractionally quantized for boson transport but zero for electron transport in the absence of interactions \cite{PhysRevLett.104.170601,goswami2016geometric,PhysRevB.100.245416,PhysRevResearch.5.043045}. Higher order fluctuations are however nonzero for both electronic and bosonic transport \cite{goswami2016geometric,giri2019nonequilibrium}. 

The geometric cumulant generating function doesn't obey Gallavotti-Cohen type of linear shift symmetry, \cite{gallavotti1995dynamical,esposito2009nonequilibrium,jarzynski2004classical} reservoiring to violation of steady state fluctuation theorems and thermodynamic uncertainty relationships \cite{PhysRevLett.104.170601,goswami2016geometric,PhysRevE.96.052129,giri2019nonequilibrium}. Such violations triggered the necessity of including geometric correction factors in fluctuation theorems and uncertainty relationships \cite{lu2022geometric,wang2022geometric,PhysRevE.107.024135,PhysRevE.102.012115}. In addition, violation of universal behaviour of efficiency at maximum power, and failure to identify particle exchange probability distributions  using large deviation theories in heat engines and thermoelectric devices are also reported \cite{PhysRevE.96.052129,giri2019nonequilibrium} along with geometric bounds on power \cite{eglinton2022geometric}. It has also been shown that the presence of interactions between electrons is necessary to facilitate geometric contributions to the flux, as demonstrated in an interacting double quantum dot \cite{yuge2012geometrical}. Driving the chemical potentials of the two electronic reservoirs in such a system was reported to yield non-zero geometric fluxes which increase linearly as a function of the interaction energy \cite{yuge2012geometrical}.

As claimed by Yuge et al \cite{yuge2012geometrical}, although not explicitly, interaction effects  play a role in the observation of finite geometricities in nonequilibrium spin-resolved electronic transport. In this work, we focus on such an interacting quantum electronic junction where the statistics can be studied in the spin-resolved basis, i.e. a nonequilibrium spintronic junction. Both attractive and repulsive interactions between electrons of the system are taken into account. Such interactions allow the possibility of two spins simultaneously getting transported across the junction, which is usually referred to as the cotunneling of electrons \cite{golovach2004transport,jiang2012inelastic,rudge2018distribution,schinabeck2020hierarchical,carmi2012enhanced,kaasbjerg2015full,xue2019non,han2010nonequilibrium,XUE201939}. This is the subject behind the letter: to address the influence of cotunneling in the geometric contributions during electron exchange statistics in the spin-resolved basis. The dynamics of such nonequilibrium cotunneling electronic systems have already been studied from several perspectives with the master equation framework being an elegant approach \cite{rudge2018distribution,schinabeck2020hierarchical,carmi2012enhanced,XUE201939}.

In this work, we investigate how the two different processes, sequential and inelastic cotunneling of electrons influence the geometric contributions to the spin-resolved electron transport statistics. Earlier studies focused only on the role of interaction energy on the geometric flux of a coupled double quantum dot. Here, we systematically use the information obtained from an analytically derived quantum master equation for a spintronic junction where the population and the Fock-space coherences are coupled. Using the well-established full counting statistics (FCS) formalism \cite{esposito2009nonequilibrium,harbola2007statistics,RevModPhys.83.771}, we perform a thorough study of the geometric flux. We also estimate the entropy production using a recently proposed thermodynamic uncertainty relationship \cite{lu2022geometric}.

The work is organized as follows. Firstly, we introduce the spintronic junction and present the quantum master equation for the reduced system dynamics within a full counting statistical method by including both sequential and cotunneling components. After that, we evaluate the geometrical statistics in the presence of cotunneling under three separate system-reservoirs coupling scenarios. Finally, we show how to use the geometric uncertainty relationship to estimate the entropy of the junction. 
\section{Formalism and Model}
\begin{figure}
    \centering
    \includegraphics[width =8.5cm]{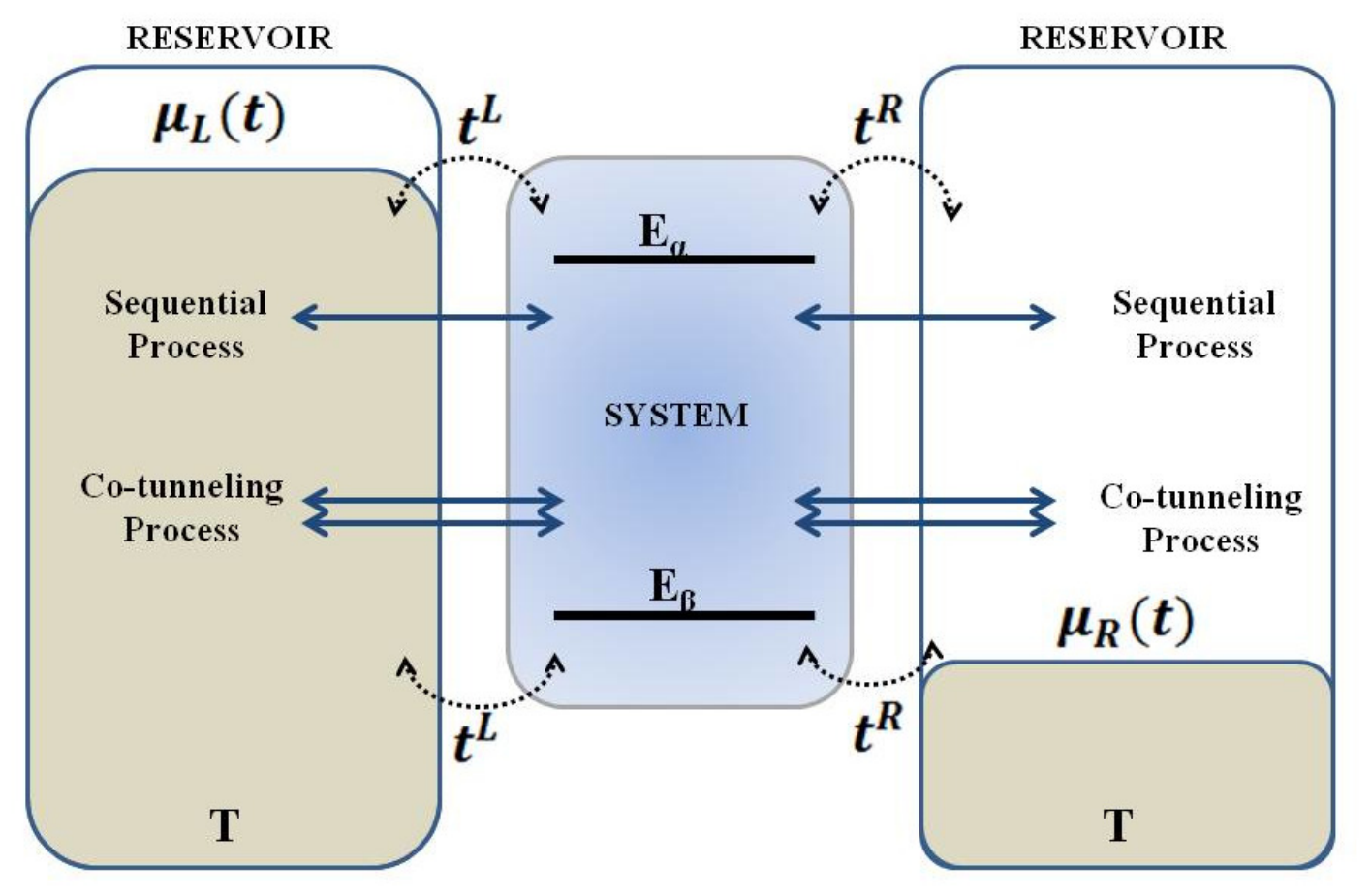}
    \caption{ Schematic illustration of the spintronic quantum junction model with two fermionic reservoirs each at a temperature $T$. The  spin orbitals' energies are $E_{\alpha}$ and $E_{\beta}$.  The chemical potentials for left and right reservoirs are time-dependent  $\mu_{_{L}}(t)$ and $\mu_{_{R}}(t)$ respectively due to an external phase different driving protocol. The solid arrows represent the different possible transitions between the reservoirs and the system. The tunnelling amplitude/coupling strength between the reservoir and the systems is denoted by $t_{L}$ and $t_{R}$.}
    \label{fig:schematic}
\end{figure}
We consider a spintronic junction with two spin orbitals coupled to two electronic reservoirs, the schematic of which is shown in Fig.(\ref{fig:schematic}). The spinless version of this is junction is analogous to a double quantum dot coupled to two electronic reservoirs \cite{golovach2004transport,rudge2018distribution,carmi2012enhanced}. The Hamiltonian of the spintronic junction has four additive components:  a system Hamiltonian, $\hat H_s$, a left (right) side reservoir Hamiltonian, $\hat H_{L(R)}$, with time-dependent chemical potential $\mu_{_L}(t) (\mu_{_R}(t))$, and temperature $T_{L}(T_{R})$  and a weak system-bath coupled Hamiltonian, $\hat H_{sb}$. The total Hamiltonian is $\hat H_o+\hat H_L+\hat H_R +\hat H_{sb}$.  
The reservoirs are modelled as a sea of non-interacting electrons whose Hamiltonian is of the form,
\begin{align}
    \hat{H}_{L(R)}(t) = \displaystyle\sum_{a\in L(R)}\sum_{\sigma=\alpha,\beta}(\varepsilon_{a,\sigma}-\mu_{a,\sigma}^{}(t,\phi))\hat{c}^{\dagger}_{a,\sigma}\hat{c}_{a,\sigma}
\end{align}
The operators $\hat{c}^{\dagger}_{a,\sigma}$($\hat{c}_{a,\sigma}$) represent the creation (annihilation) of an electron in the single-particle reservoir $a \in L,R$ in either $\alpha$ or $\beta$ spin state with energy $\varepsilon_{a,\sigma}$. The density matrix of the reservoirs is assumed to be diagonal whose chemical potentials are being driven in time with a constant phase difference $\phi$. We assume that this driving doesn't affect the internal dynamics of the bath, i.e. the external driving is adiabatic and the system and reservoirs have well-separated timescales. 
The system Hamiltonian, $\hat H_o = \hat H_s+\hat H_U$, given by,
\begin{align}
    \hat{H}_s = \displaystyle\sum_{\sigma\in\alpha,\beta} E_{\sigma} \hat{c}^{\dagger}_{\sigma} \hat{c}_{\sigma}, ~ \hat{H}_{U} = U \hat{c}_{\sigma}^\dagger \hat{c}_{\sigma} \hat{c}_{\sigma'}^\dagger \hat{c}_{\sigma'}, \sigma \in\alpha,\beta 
\end{align}
with the onsite Hamiltonian, $\hat H_s$ whose system operator $\hat c_\sigma^{\dag}$($\hat c_\sigma$) creates (annihilates) an electron with spin $\sigma$ on the spintronic level with energy $E_\sigma$. $\hat H_U$ is the onsite diagonalized interaction Hamiltonian that accounts for the attractive ($U<0$) and repulsive ($U>0$) interactions between two spins, $\alpha$ and $\beta$.  
The coupling Hamiltonian between the system and the reservoir is, 
\begin{align}
\hat{H}_{sb} = t_{L}\displaystyle\sum_{a\in L}\sum_{\sigma\in\alpha,\beta}\hat{c}^{\dagger}_{a,\sigma}\hat{c}_{a}+ t_{R}\displaystyle\sum_{a\in R}\sum_{\sigma\in\alpha,\beta}\hat{c}^{\dagger}_{a,\sigma}\hat{c}_{a} +h.c
\end{align}
where $t_{L(R)}$ is used to denote the tunnelling amplitude between the electron and corresponding left (right) reservoirs. 
In the weak coupling limit, interactions due to the system-bath coupling can be captured up to the second order by deriving an adiabatic quantum master equation for the reduced system density matrix
\cite{hino2020fluctuation,njp-adia,PhysRevE.106.024131,yuge2012geometrical}. The reduced system density matrix in the spin-resolved Fock state basis is \cite{harbola2006quantum,rudge2018distribution,carmi2012enhanced} $|\rho\rangle= (\rho_{{0,0}},\rho_{{\alpha,\alpha}},\rho_{{\beta,\beta}},\rho_{{2,2}},\rho_{{\alpha,\beta}},\rho_{{\beta,\alpha}})$.  Here, $\rho_{{n,n}}, n =0,\alpha$ and $\beta$ are the populations corresponding to the unoccupied state $|0\rangle$, $\alpha$ spin occupied spin-orbital $|\alpha\rangle$ and $\beta$ spin occupied spin-orbital $|\beta\rangle$. $\rho_{2,2}$ is the double occupied state which is not spin-resolved. 
$\rho_{\alpha\beta} (\rho_{\beta,\alpha})$ are the Fock space coherences between the singly occupied spin resolved states. 
 The dynamics of the system, 
within the adiabatic evolution limit can be written as
$|\rho(t)\rangle=\breve {\cal L}(t)|\rho\rangle$, with the superoperator Liouvillian being,
\begin{widetext}
\begin{align}\label{6x6 rate}
\breve {\cal L}(t)&=\begin{bmatrix}
-\sum_{k\not=0} \omega_{0,k} & \omega_{\alpha,0} & \omega_{\beta,0} & \omega_{2,0} & (\omega^{\beta,0}_{\alpha,0}+\omega^{\alpha,0}_{\beta,0})& (\omega^{\alpha,0}_{\beta,0}+\omega^{\beta,0}_{\alpha,0}) \\
\omega_{0,\alpha} & -\sum_{k\not=\alpha} \omega_{\alpha,k} & \omega_{\beta,\alpha} & \omega_{2,\alpha} & (-\omega^{\alpha,0}_{\beta,0} + \omega^{0,\alpha}_{0,\beta}) & (-\omega^{\alpha,0}_{\beta,0}-\omega^{0,\alpha}_{0,\beta}) \\
\omega_{0,\beta} &  \omega_{\alpha,\beta} & -\sum_{k\not=\beta} \omega_{\beta,k} & \omega_{2,\beta} & (-\omega^{\beta,0}_{\alpha,0} + \omega^{0,\beta}_{0,\alpha} ) & (-\omega^{\beta,0}_{\alpha,0} + \omega^{0,\beta}_{0,\alpha}) \\
\omega_{0,2} &  \omega_{\alpha,2} & \omega_{\beta,2} & -\sum_{k\not=2} \omega_{2,k} & (- \omega^{0,\alpha}_{0,\beta} -\omega^{0,\beta}_{0,\alpha}) &( -\omega^{0,\alpha}_{0,\beta} - \omega^{0,\beta}_{0,\alpha}) \\
(\omega^{0,\beta}_{0,\alpha} + \omega^{0,\alpha}_{0,\beta}) & (-\omega^{\beta,0}_{\alpha,0}+\omega^{0,\beta}_{0,\alpha}) & (-\omega^{\alpha,0}_{\beta,0} + \omega^{0,\alpha}_{0,\beta})& (-\omega^{\alpha,0}_{\beta,0} - \omega^{\beta,0}_{\alpha,0}) & -\chi & 0 \\
(\omega^{0,\alpha}_{0,\beta} + \omega^{0,\beta}_{0,\alpha}) & (-\omega^{\beta,0}_{\alpha,0}+\omega^{0,\beta}_{0,\alpha}) & (-\omega^{\alpha,0}_{\beta,0}+\omega^{0,\alpha}_{0,\beta}) & (-\omega^{\beta,0}_{\alpha,0} - \omega^{\alpha,0}_{\beta,0} ) & 0 & -\chi
\end{bmatrix},\\
\chi &= \omega_{\beta,0} + \omega_{0,\beta} + \omega_{\alpha,0} + \omega_{0,\alpha}
\end{align}
\end{widetext}
In this adiabatic master equation for the spintronic junction, all the matrix elements in the superoperator ${\cal L}(t)$ are time-dependent quantities, i.e., $\omega^{(\cdots)}_{i,f}:= \omega^{(\cdots)}_{i,f}(t)$. The source of the time dependence is introduced by externally driving the chemical potential of the reservoirs' $\mu_L$ and $\mu_R$ in a periodic manner albeit in an adiabatic fashion \cite{sinitsyn2007berry}. 
We omit writing of the explicit time-dependent argument on each term for brevity. 
Each matrix element of the superoperator in Eq.(\ref{6x6 rate}) can be interpreted as a  net rate term, $\omega_{i,f}=\sum_{(\cdots)}\omega^{(\cdots)}_{i,f}$ which is the total rate of electron transfer from state $|i\rangle$ to $|f\rangle$. The individual rates are denoted as ${{\omega}^{(\cdots)}_{i,f}}$, where $(\cdots) \in \rightarrow,\leftarrow,\leftrightarrow,\Rightarrow,\Leftarrow,\leftrightarrows$ or $\rightleftarrows$, labelled to distinguish different possible tunnelling or cotunneling processes involving the left or right reservoirs with the spin-orbitals of the system \cite{carmi2012enhanced}. Each term, $\omega^{(\cdots)}_{i,f}$ has its own physical significance.
For example, let us consider a transition from state $|0\rangle$ to $|\alpha\rangle$. This happens when an electron is tunnelling from the left reservoir or the right reservoir into the level $E_{\alpha}$ while the level $E_{\beta}$ is empty. The rate of tunnelling from   the left to right direction involving the left reservoir and the Fock state $|\alpha\rangle$ of the system is a sequential tunnelling process with a rate denoted by$,   
    \omega^{\rightarrow}_{0,\alpha} = \Gamma_L F_{FD}(E_{\alpha} - \mu_L(t))
$. 
Similarly, for a tunneling process from the right reservoir to the Fock state $|\alpha\rangle$ of the system, is 
$
    \omega^{\leftarrow}_{0,\alpha} = \Gamma_R F_{FD}(E_{\alpha} - \mu_R(t))
$, another sequential tunneling process. This makes the total sequential tunneling rate $\omega_{0,\alpha}$ involving the Fock state $|\alpha\rangle$ and both the reservoirs to be a summation of two terms, $
    \omega_{0,\alpha} = \omega^{\leftarrow}_{0,\alpha} + \omega^{\rightarrow}_{0,\alpha}
$. We have used
$
    \Gamma_{L} = \nu \frac{2\pi}{\hbar} {|t_{L}|}^{2}, \Gamma_{R} = \nu \frac{2\pi}{\hbar} {|t_{R}|}^{2}
$. 
The Fermi function is denoted as $F_{FD}(a-x(t)) = (1 + \exp\{\beta (a-x(t))\})^{-1}$. Similarly, we can interpret the other sequential rates. 
In the fifth and the sixth rows and columns, there are matrix elements of the type $\omega^{A}_{B} = \omega^{\leftarrow}_{A} + \omega^{\rightarrow}_{B} $, where $A\neq B$ and $A,B \in (\beta,0), (\alpha,0),(0,\beta), (0,\alpha)$.  These are processes that involve the addition of two different sequential processes, $\omega^{\leftarrow}_{A}$ and $\omega^{\rightarrow}_{B}$ and allow electron transfers by coupling the population with the coherences in the system \cite{harbola2006quantum}. These terms can be eliminated from the system dynamics by decoupling the populations and the coherences using rotating wave or secular approximation \cite{harbola2006quantum, esposito2009nonequilibrium}.  
 Rates of the type $\omega^{\rightarrow}_{i,f}$ and $\omega^{\leftarrow}_{i,f}$ are the sequential rates since these involve a single electron entering or leaving the spin orbitals. Rates of the type $\omega^{(\cdots)}_{i,f}$ with $(\cdots) \in \leftrightarrows,\rightleftarrows,\Leftarrow$ or $\Rightarrow$ are called inelastic cotunneling rates. The inelastic cotunneling rates involve the exchange of two simultaneous electrons \cite{carmi2012enhanced,rudge2018distribution}. For example, consider the process $\omega_{\alpha,\beta}$ i.e., the transition from $|\alpha\rangle$ to $|\beta\rangle$ state. This process is a sum of four different processes, $\omega^{\Leftarrow}_{\alpha,\beta} + \omega^{\Rightarrow}_{\alpha,\beta} + \omega^{\rightleftarrows}_{\alpha,\beta} + \omega^{\leftrightarrows}_{\alpha,\beta}$. The first term (second term) $\omega^{\Leftarrow}_{\alpha,\beta} (\omega^{\Rightarrow}_{\alpha,\beta})$ describes the process where one electron has moved out from level $E_{\alpha}$ to the left reservoir (right reservoir) and simultaneously an electron has entered the level $E_{\beta}$ from the right reservoir (left reservoir). The third term (fourth term), $\omega^{\rightleftarrows}_{\alpha,\beta} (\omega^{\leftrightarrows}_{\alpha,\beta})$ describes the process where one electron moves out to the right reservoir (left reservoir) from level $E_{\alpha}$ and simultaneously one electron enters into level $E_{\beta}$ from the right reservoir (left reservoir). 

Lastly, we describe the process $\omega_{2,0}$, which is a sum of three different rates $\omega^{\Leftarrow}_{2,0} + \omega^{\Rightarrow}_{2,0} + \omega^{\leftrightarrow}_{2,0}$. The first term (second term) $\omega^{\Leftarrow}_{2,0}(\omega^{\Rightarrow}_{2,0})$ describes the rate which involves simultaneous leaving of the two electrons from level $E_{\alpha}$ and $E_{\beta}$ to the left reservoir(right reservoir) which was previously a doubly occupied Fock state. 
 $\omega^{\leftrightarrow}_{2,0}$ describes the process in which two electrons leave the two levels by simultaneous tunneling of one to the right and the other to the left or vice versa.
 There also exists elastic cotunneling rates \cite{carmi2012enhanced}. We do not take these into account since electron transfer involving such rates have intermediate processes with Fock states that differ by at least two electrons. These can be taken into account by including Fock space coherences between states with different number of particles. The mathematical expression for all the sequential rates can be obtained analytically \cite{harbola2006quantum}. The cotunneling rates can also be obtained analytically following a regularization procedure \cite{carmi2012enhanced,rudge2018distribution} in the equal temperature setting. All the mathematical expressions of the rates are given in the appendix.

Within the full counting statistical framework \cite{esposito2009nonequilibrium,harbola2007statistics}, each electron exchange process can be tracked via the auxiliary counting vector, ${\boldsymbol \zeta}$ and a twisted generator of the type $|\dot\rho ({\boldsymbol \zeta})\rangle = \breve{\cal L}(\boldsymbol \zeta,t)|\rho\rangle$.
The counting field vector ${\boldsymbol \zeta}$ contains the individual tracking parameters $\zeta_{if}^{(\cdots)}$ that track the frequency $\omega_{if}^{(\cdots)}$ and allows one to estimate the statistical moments and cumulants of electronic transport associated in each process in the junction. Usually, this counting process is carried out at the steady-state, where the largest eigenvalue of $\breve{\cal{L}}(\boldsymbol \zeta,t)$ represents a cumulant generating function, $S(\boldsymbol \zeta)$. The $k-$th cumulant is obtained by evaluating the $k-$th derivative of $S(\boldsymbol \zeta)$ and by setting $\boldsymbol \zeta=0$ \cite{esposito2009nonequilibrium,jarzynski2004classical,nazarov2003full,kaasbjerg2015full}. 
In the present case, we do not focus on the individual statistics of each process but on the electronic exchange between the system and the right reservoir. Hence all terms in ${\boldsymbol \zeta}$ that track the spin exchanges in the left side of the junction are taken to be zero. Further, since we focus on the exchange process, all counting fields that track the electrons transferred from the system to the right reservoir are opposite in sign to the counting fields that track the electrons transferred to the system from the right reservoir \cite{esposito2009nonequilibrium,harbola2006quantum}. Further, the processes that involve two simultaneous electron exchanges due to cotunneling are scaled twofold. Thus the entire counting vector $ \boldsymbol \zeta$ reduces to a single counting parameter $\zeta$. The generating vector that counts the net number of electrons exchanged at the right terminal, $\breve{\cal L}(\zeta,t)$ obtained from Eq.(\ref{6x6 rate}) has the following $\zeta$-dependent terms,  
\begin{align}
\label{eq:5}
    \omega_{0,\alpha} &= \omega^{\leftarrow}_{0,\alpha}e^{\zeta} + \omega^{\rightarrow}_{0,\alpha} \\
    \omega_{\alpha,0} &= \omega^{\leftarrow}_{\alpha,0} + \omega^{\rightarrow}_{\alpha,0}e^{-\zeta} \\
    \omega_{0,\beta} &= \omega^{\leftarrow}_{0,\beta}e^{\zeta} + \omega^{\rightarrow}_{0,\beta} \\
    \omega_{\beta,0} &= \omega^{\leftarrow}_{\beta,0} + \omega^{\rightarrow}_{\beta,0}e^{-\zeta} \\
    \omega_{\alpha,\beta} &= \omega^{\Leftarrow}_{\alpha,\beta}e^{\zeta} + \omega^{\Rightarrow}_{\alpha,\beta}e^{-\zeta} + \omega^{\rightleftarrows}_{\alpha,\beta} + \omega^{\leftrightarrows}_{\alpha,\beta} \\
    \omega_{\beta,\alpha} &= \omega^{\Leftarrow}_{\beta,\alpha}e^{\zeta} + \omega^{\Rightarrow}_{\beta,\alpha}e^{-\zeta} + \omega^{\rightleftarrows}_{\beta,\alpha} + \omega^{\leftrightarrows}_{\beta,\alpha} \\
    \omega_{\alpha,2} &= \omega^{\leftarrow}_{\alpha,2}e^{\zeta} + \omega^{\rightarrow}_{\alpha,2} \\
    \omega_{2,\alpha} &= \omega^{\leftarrow}_{2,\alpha} + \omega^{\rightarrow}_{2,\alpha}e^{-\zeta} \\
    \omega_{\beta,2} &= \omega^{\leftarrow}_{\beta,2}e^{\zeta} + \omega^{\rightarrow}_{\beta,2} \\
    \omega_{2,\beta} &= \omega^{\leftarrow}_{2,\beta} + \omega^{\rightarrow}_{2,\beta}e^{-\zeta} \\
    \omega_{0,2} &= \omega^{\Leftarrow}_{0,2}e^{2\zeta} + \omega^{\Rightarrow}_{0,2} + \omega^{\leftrightarrow}_{0,2}e^{\zeta} \\
    \omega_{2,0} &= \omega^{\Leftarrow}_{2,0} + \omega^{\Rightarrow}_{2,0}e^{-2\zeta} + \omega^{\leftrightarrow}_{2,0}e^{-\zeta} \\
    \label{eq-hybrid-rates}
    \omega^{0,\alpha}_{0,\beta} &= \omega^{\leftarrow}_{0,\alpha}e^{\zeta} + \omega^{\rightarrow}_{0,\beta} \\
    \omega^{0,\beta}_{0,\alpha} &= \omega^{\leftarrow}_{0,\beta}e^{\zeta} + \omega^{\rightarrow}_{0,\alpha} \\
    \omega^{\alpha,0}_{\beta,0} &= \omega^{\leftarrow}_{\alpha,0} + \omega^{\rightarrow}_{\beta,0}e^{-\zeta} \\
    \label{eq-hybrid-rates-}
    \omega^{\beta,0}_{\alpha,0} &= \omega^{\leftarrow}_{\beta,0} + \omega^{\rightarrow}_{\alpha,0}e^{-\zeta},
\end{align}
which allows full tracking of the total electronic exchanges in the presence of both sequential and cotunneling processes. In the secular limit, the fifth and sixth, rows and columns of Eq.(\ref{6x6 rate}) are absent and the density matrix is reduced to a vector containing only the populations of the four Fock states. In this scenario, the hybrid rates in  Eq.(\ref{eq-hybrid-rates} -\ref{eq-hybrid-rates-}), do not contribute to the steady-state statistics and the dimension of the Liouvillian, Eq.({\ref{6x6 rate}}) reduces to $4\times4$.

\section{Geometric Statistics}
  
The geometricity or the geometric phase-like effects in a nonequilibrium quantum junction have its roots embedded in the scaled cumulant generating function which keeps track of the number of electrons exchanged between the system and the reservoir. The total scaled cumulant generating function is additively separable into two parts, $S(\zeta)=S_d(\zeta)+S_g(\zeta)$, where the dynamic part is $S_{d}(\zeta)$ and the geometric part is $S_{g}({\zeta})$ \cite{sinitsyn2007berry,sinitsyn2009stochastic,goswami2016geometric,PhysRevLett.104.170601}. The geometric cumulant generating function has a general expression,
\begin{align}
\label{eq:geometric part = line intergal time}
    S_{g}{({\zeta})} &= \frac{-1}{t_p} \int_{0}^{t_p} \langle L_{\circ}(\zeta,t)|\frac{\partial}{\partial t}|R_{\circ}(\zeta,t)\rangle dt
\end{align}
\\
where $|R_{\circ}(\zeta,t)\rangle$ and $|L_{\circ}(\zeta,t)\rangle$ represents the instantaneous right and left eigenvectors at the time ${t}$ for ${\zeta}$ dependent Liouvillian superoperator corresponding to the largest instantaneous positive eigenvalue $\lambda_{\circ}(\zeta,t)$. If we define a surface $S$ by a closed contour in the parameter space involving two parameters ${\mu_{_L}}$ and ${\mu_{_R}}$, then Eq.(\ref{eq:geometric part = line intergal time}) can be further transformed via using the Stokes formula as $
    S_g(\zeta) = -\frac{1}{t_p} {\oiint}_{S} B^{\zeta}(\mu_{_L},\mu_{_R}) d\mu_{_L} d\mu_{_R} 
$.
The integrand $B^{\zeta}(\mu_{_L}\mu_{_R})$ is analogous to the gauge-invariant Pancharatnam-Berry curvature or geometric curvature \cite{sinitsyn2007berry,PhysRevLett.104.170601,goswami2016geometric}. The geometric statistics are quantified by evaluating the geometric cumulants,
\begin{align}
\label{eq:jkg}
 j^{(k)}_g=\frac{d^k}{d\zeta^k}S_g(\zeta)|_{\zeta=0}
\end{align}
where $j^{(k)}_g$ is the $k$-th geometric cumulant. When $k=1(2)$, we obtain the geometric flux (noise). Depending on the phase difference $\phi$, both these cumulants can be positive or negative. Note that the total noise (sum of dynamic and geometric) is always positive but the geometric noise can be negative but is never greater than the dynamic noise \cite{sinitsyn2009stochastic}. The dynamic cumulants can be obtained by using $j^{(k)}_d=\frac{d^k}{d\zeta^k}S_d(\zeta)|_{\zeta=0}$ and their behavior is akin to what is seen for non-driven case, albeit with scaling factors. We do not discuss these in this work. In the next section, we present our results for the geometric cumulants. 

\section{Results and Discussion}
\begin{figure}
    \centering
    \includegraphics[width =8.5cm]{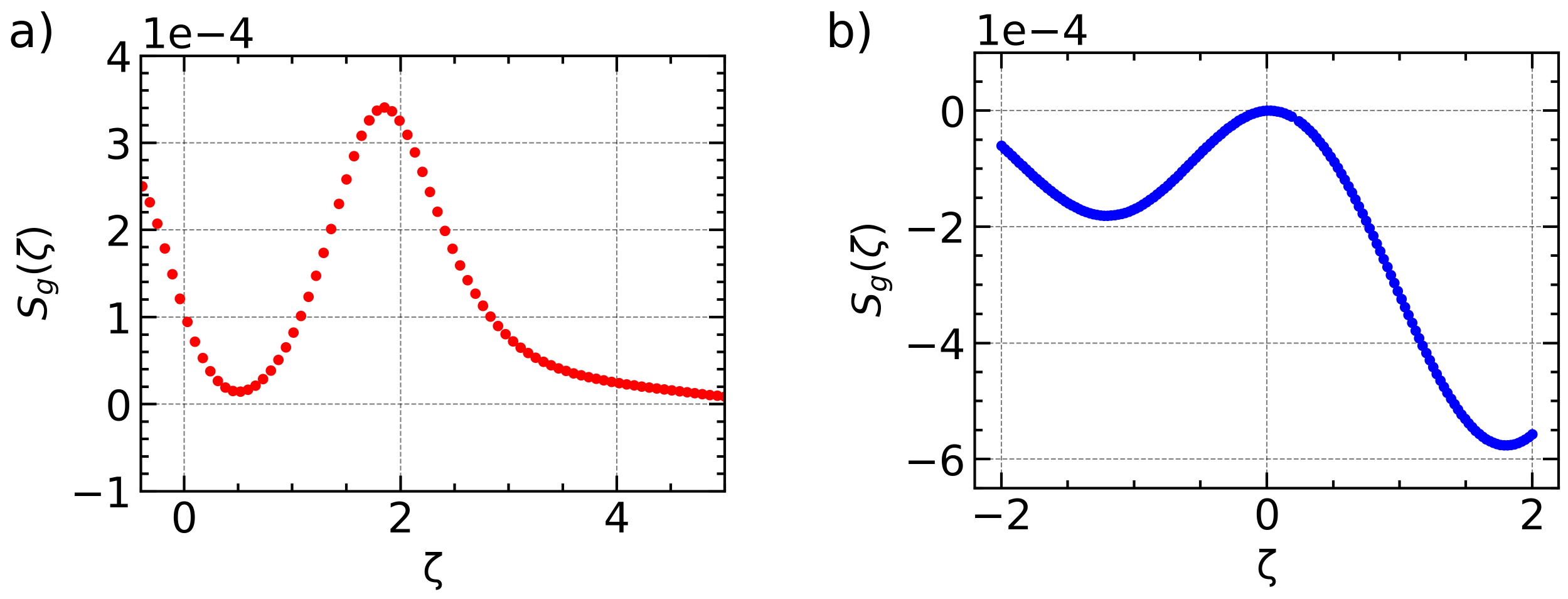}
    \caption{Behavior of the geometric cumulant generating function in the presence of cotunneling, $S_{g}(\zeta)$ obtained using Eq.(\ref{6x6 rate}) and Eq.(\ref{eq:geometric part = line intergal time}) for both (a) non-secular and (b) secular case vs the auxiliary counting field $\zeta$  with the following parameters for $t_{L} \approx t_{R}$: $[\Gamma = 2(\Gamma_{L} + \Gamma_{R})]$ : $K_{B}T = \hbar\Gamma, E_{\alpha} = 3 meV, E_{\beta} = E_{\alpha} - \hbar\Gamma, U = 1.5meV, \mu^{\circ}_{L} = 4meV, eV_{bias} = 1.5\hbar\Gamma, m = 0.09$ and $\phi = \frac{\pi}{3}$. These parameters are kept fixed throughout the manuscript unless stated otherwise.}
    \label{fig:GP tl approx tr}
\end{figure}
\begin{figure}
    \centering 
    \includegraphics[width=8.5cm]{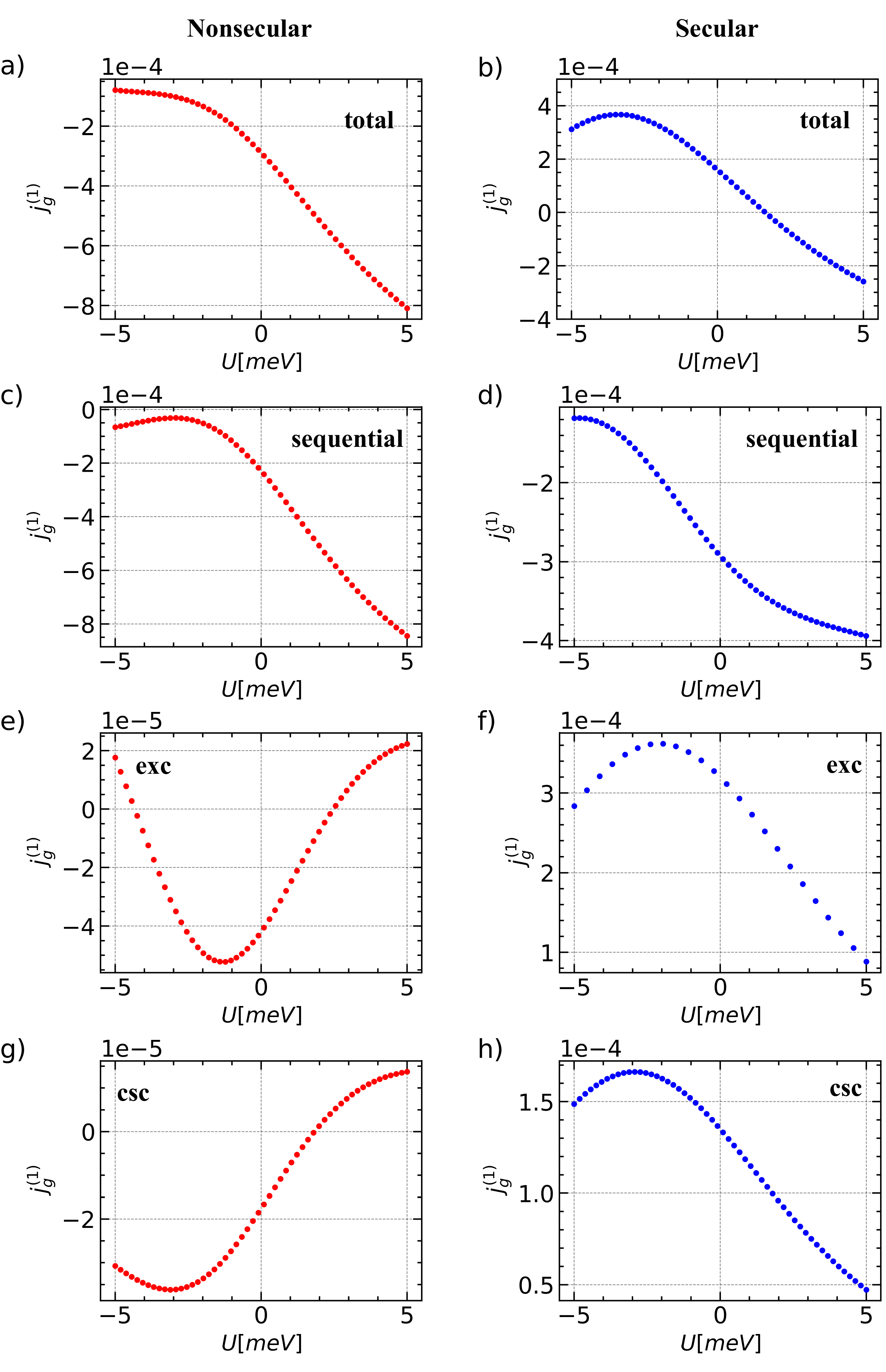}
    \caption{(Color Online) Behavior of the total geometric flux, $j^{(1)}_g$  (from Eq.(\ref{eq:jkg}) using Eq.(\ref{6x6 rate}) with Eqs. (\ref{eq:5}-\ref{eq-hybrid-rates-}) for $k=1$ and $t_{L}\approx t_{R}$) as a function of the interaction potential $U$, (a) in the presence  and (b) absence of coherence. 
    Plots in red (left side) and blue (right side) correspond to the non secular and secular case respectively. 
    Contribution to the geometric flux from only the sequential electron transfer process (from Eq.(\ref{6x6 rate-seq})), in presence (c) and absence (d) of coherences. Contribution only from the exchange cotunneling processes (from Eq.(\ref{6x6 rate-exc})) in presence (e) and (f) absence of coherences. Contribution only from the doubly charge-separated processes (from Eq.(\ref{6x6 rate double electron tracking})) are shown in (g) with coherences and (h) without coherences.}
    \label{fig:j1g tl approx tr}
\end{figure}
In our calculations, we take the externally controlled adiabatic driving to be of the following form, $\mu_L(t) = \mu_L^o(1 - m^2\sin^2({\Omega}t))$ and $\mu_R(t) = \mu_R^o(1 - m^2\sin^2({\Omega}t+\phi))$, with $m$ being the amplitude of the externally controlled driving protocol with initial values $\mu_L^o$ and $\mu_R^o$. We obtain $S_g(\zeta)$ from Eq.(\ref{eq:geometric part = line intergal time}) by numerically evaluating the left and right eigenvectors of Eq.(\ref{6x6 rate}). We do this for two cases: for Eq.(\ref{6x6 rate}) (plotted red) and under the rotating wave or secular approximation  (plotted blue) and using Eqs. (6-21), where the fifth and sixth rows and columns of Eq.(\ref{6x6 rate}) are absent. The cumulant generating functions are shown in Fig.(\ref{fig:GP tl approx tr}a and \ref{fig:GP tl approx tr}b) for both nonsecular and secular cases with $t_L\approx t_R$.

In Fig.(\ref{fig:j1g tl approx tr}(a),(b)), we show how the total geometric flux, $j^{(1)}_{g}$ varies as a function of the interaction $U$ for the two cases with $t_L\approx t_R$ using Eq.(\ref{6x6 rate}) with Eq.(\ref{eq:geometric part = line intergal time}). For the non-secular case (plotted red), the net geometric flux, Fig.(\ref{fig:j1g tl approx tr}a) is negative and nonlinear as a function of $U$. For repulsive coulomb interaction $U>0$, a nonlinear dependence on $U$ is seen. Linear dependence has been previously reported in interacting double quantum dots\cite{takahashi2020full}. For attractive interactions between the spins $U<0$, the net geometric flux is a nonlinear function of $U$.
Whereas, for the secular case (plotted blue), the net geometric flux,Fig.(\ref{fig:j1g tl approx tr}b) changes from negative to positive as $U$ changes from being attractive to repulsive. It is also nonlinear in the attractive regime ($U<0$) and almost linear in the repulsive regime ($U>0$). In addition,a maxima is observed at $U=-3.5$ whereas such maxima is not observed in the non-secular case, Fig.(\ref{fig:j1g tl approx tr}a). 

The evaluated $j^{(1)}_g$ using Eq.(\ref{6x6 rate}) with Eq.(\ref{eq:geometric part = line intergal time}) along with Eqs.(6-21) is the total geometric flux between the system and the right reservoir. It has contributions from both the sequential and cotunneling processes. We can identify the contributions from the sequential and cotunneling processes as follows. There are rates of the type $\omega_{if}^{(\cdots)}$, where $(\cdots)\in\leftarrow$ or $\rightarrow$ and change of states between $|i\rangle$ and $|f\rangle$ differ only by a single electron. These are sequential processes and can be separately tracked by appropriately identifying the counting field $\boldsymbol{\zeta}$ in Eq.(\ref{6x6 rate}). In this case, the twisted generator, $\breve{\cal{L}}_{seq}(\zeta,t)$ involves only the sequential processes at the right reservoir and is given by ,
\begin{widetext}
\begin{align}\label{6x6 rate-seq}
\breve{\cal{L}}_{seq}(\zeta,t)=
\begin{bmatrix}
-\sum_{k\not=0}\omega_{0,k} & \descbox{\omega_{\alpha,0}} & \descbox{\omega_{\beta,0}} & {\omega_{2,0}} & \descbox{(\omega^{\beta,0}_{\alpha,0}+\omega^{\alpha,0}_{\beta,0})}& \descbox{(\omega^{\alpha,0}_{\beta,0}+\omega^{\beta,0}_{\alpha,0})} \\
\descbox{\omega_{0,\alpha}} & {-\sum_{k\not=\alpha} \omega_{\alpha,k}} & {\omega_{\beta,\alpha}} & \descbox{\omega_{2,\alpha}} & \descbox{(-\omega^{\alpha,0}_{\beta,0} + \omega^{0,\alpha}_{0,\beta})} & \descbox{(-\omega^{\alpha,0}_{\beta,0}-\omega^{0,\alpha}_{0,\beta})} \\
\descbox{\omega_{0,\beta}} &  {\omega_{\alpha,\beta}} & -\sum_{k\not=\beta} \omega_{\beta,k} & \descbox{\omega_{2,\beta}} & \descbox{(-\omega^{\beta,0}_{\alpha,0} + \omega^{0,\beta}_{0,\alpha})} & \descbox{(-\omega^{\beta,0}_{\alpha,0} + \omega^{0,\beta}_{0,\alpha})} \\
{\omega_{0,2}} &  \descbox{\omega_{\alpha,2}} & \descbox{\omega_{\beta,2}} & -\sum_{k\not=2} \omega_{2,k} & \descbox{(- \omega^{0,\alpha}_{0,\beta} -\omega^{0,\beta}_{0,\alpha})} & \descbox{( -\omega^{0,\alpha}_{0,\beta} - \omega^{0,\beta}_{0,\alpha})} \\
\descbox{(\omega^{0,\beta}_{0,\alpha} + \omega^{0,\alpha}_{0,\beta})} & \descbox{(-\omega^{\beta,0}_{\alpha,0}+\omega^{0,\beta}_{0,\alpha})} & \descbox{(-\omega^{\alpha,0}_{\beta,0} + \omega^{0,\alpha}_{0,\beta})}& \descbox{(-\omega^{\alpha,0}_{\beta,0} - \omega^{\beta,0}_{\alpha,0})} & -\chi & 0 \\
\descbox{(\omega^{0,\alpha}_{0,\beta} + \omega^{0,\beta}_{0,\alpha})} & \descbox{(-\omega^{\beta,0}_{\alpha,0}+\omega^{0,\beta}_{0,\alpha})} & \descbox{(-\omega^{\alpha,0}_{\beta,0}+\omega^{0,\alpha}_{0,\beta})} & \descbox{(-\omega^{\beta,0}_{\alpha,0} - \omega^{\alpha,0}_{\beta,0})} & 0 & -\chi
\end{bmatrix}
\end{align}
\end{widetext}
where the matrix elements highlighted by the boxes indicate the terms where the counting field $\zeta$ is present. Explicitly, in Eq.(\ref{6x6 rate-seq}), we have the counting field dependencies in the following rates,

\begin{align}
    \omega_{0,\alpha} &= \omega^{\leftarrow}_{0,\alpha}e^{\zeta} + \omega^{\rightarrow}_{0,\alpha} \\
    \omega_{\alpha,0} &= \omega^{\leftarrow}_{\alpha,0} + \omega^{\rightarrow}_{\alpha,0}e^{-\zeta} \\
    \omega_{0,\beta} &= \omega^{\leftarrow}_{0,\beta}e^{\zeta} + \omega^{\rightarrow}_{0,\beta} \\
    \omega_{\beta,0} &= \omega^{\leftarrow}_{\beta,0} + \omega^{\rightarrow}_{\beta,0}e^{-\zeta} \\
    \omega_{\alpha,2} &= \omega^{\leftarrow}_{\alpha,2}e^{\zeta} + \omega^{\rightarrow}_{\alpha,2} \\
    \omega_{2,\alpha} &= \omega^{\leftarrow}_{2,\alpha} + \omega^{\rightarrow}_{2,\alpha}e^{-\zeta} \\ 
    \omega_{\beta,2} &= \omega^{\leftarrow}_{\beta,2}e^{\zeta} + \omega^{\rightarrow}_{\beta,2} \\
    \omega_{2,\beta} &= \omega^{\leftarrow}_{2,\beta} + \omega^{\rightarrow}_{2,\beta}e^{-\zeta}
\end{align}
\begin{align}
    \omega^{0,\alpha}_{0,\beta} &= \omega^{\leftarrow}_{0,\alpha}e^{\zeta} + \omega^{\rightarrow}_{0,\beta} \\
    \omega^{0,\beta}_{0,\alpha} &= \omega^{\leftarrow}_{0,\beta}e^{\zeta} + \omega^{\rightarrow}_{0,\alpha} \\
    \omega^{\alpha,0}_{\beta,0} &= \omega^{\leftarrow}_{\alpha,0} + \omega^{\rightarrow}_{\beta,0}e^{-\zeta} \\
    \omega^{\beta,0}_{\alpha,0} &= \omega^{\leftarrow}_{\beta,0} + \omega^{\rightarrow}_{\alpha,0}e^{-\zeta}
\end{align}
As before, under secular approximation, the fifth and sixth rows and columns of Eq.(\ref{6x6 rate-seq}) do not contribute to the statistics. We numerically obtain the contribution due to the sequential processes to the total flux, by evaluating the largest eigenvalue and its corresponding left and right eigenvectors of Eq. (\ref{6x6 rate-seq}). We then use these eigenvectors in Eq.(\ref{eq:geometric part = line intergal time}) to obtain the contribution to $j^{(1)}_g$ from the sequential processes which is shown in Fig.(\ref{fig:j1g tl approx tr}c and \ref{fig:j1g tl approx tr}d) for the non-secular and secular case respectively. Under both cases, the sequential fluxes are negative. For the nonsecular case, the sequential flux (Fig.\ref{fig:j1g tl approx tr}c) is almost identical to the total flux (Fig.\ref{fig:j1g tl approx tr}a).  This indicates that the sequential processes contribute more to the geometric flux in the presence of coherences than in the absence of coherences. Thus, the effect of cotunneling processes is minimal when the coherences and populations are coupled. However, for the secular case, we expect contributions from the cotunneling electrons since the flux behavior is different as shown in Fig.(\ref{fig:j1g tl approx tr}f and \ref{fig:j1g tl approx tr}h). To establish this further, we separately identify the cotunneling contributions to the statistics by introducing appropriate counting fields on the cotunneling processes alone. Note that there are two different types of cotunneling processes as evident from the nature of the rate expressions.
Rates of the type $\omega_{if}^{(\cdots)}$, where $(\cdots) \in \Leftarrow, \Rightarrow, \leftrightarrows$ or $\rightleftarrows$ where transition between the states $|i\rangle$ and $|f\rangle$ occur by exchange of two electrons with both the left and right reservoirs. Effectively such rates involve the exchange of one electron with the left reservoir and one electron with the right reservoir and contributing once at the right terminal. We refer to these rates as exchange cotunneling rates.
There are also rates of the type $\omega_{if}^{(\cdots)}$, where $(\cdots) \in \Leftarrow, \Rightarrow$ or $\leftrightarrow$, and transition between the states $|i\rangle$ and $|f\rangle$ involve two electrons and both electrons are either involved with the left or the right reservoir. We refer to such cotunneling rates involving the states $|0\rangle$ and $|2\rangle$ as double charge-separated cotunneling rates. These rates contribute twice to the statistics. 
The twisted generator, $\breve{\cal L}_{exc}$ for tracking the exchange cotunneling electrons is given by
\begin{widetext}
\begin{align}\label{6x6 rate-exc}
\breve{\cal L}_{exc}=
\begin{bmatrix}
-\sum_{k\not=0}\omega_{0,k} & \omega_{\alpha,0} & \omega_{\beta,0} & \dashboxed{\omega_{2,0}} & {(\omega^{\beta,0}_{\alpha,0}+\omega^{\alpha,0}_{\beta,0})}& {(\omega^{\alpha,0}_{\beta,0}+\omega^{\beta,0}_{\alpha,0})} \\
{\omega_{0,\alpha}} & {-\sum_{k\not=\alpha} \omega_{\alpha,k}} & \descbox{\omega_{\beta,\alpha}} & {\omega_{2,\alpha}} & {(-\omega^{\alpha,0}_{\beta,0} + \omega^{0,\alpha}_{0,\beta})} & (-\omega^{\alpha,0}_{\beta,0}-\omega^{0,\alpha}_{0,\beta}) \\
{\omega_{0,\beta}} &  \descbox{\omega_{\alpha,\beta}} & -\sum_{k\not=\beta} \omega_{\beta,k} & \omega_{2,\beta} & (-\omega^{\beta,0}_{\alpha,0} + \omega^{0,\beta}_{0,\alpha} ) & (-\omega^{\beta,0}_{\alpha,0} + \omega^{0,\beta}_{0,\alpha}) \\
\dashboxed{\omega_{0,2}} &  \omega_{\alpha,2} & \omega_{\beta,2} & -\sum_{k\not=2} \omega_{2,k} & (- \omega^{0,\alpha}_{0,\beta} -\omega^{0,\beta}_{0,\alpha}) &( -\omega^{0,\alpha}_{0,\beta} - \omega^{0,\beta}_{0,\alpha}) \\
{(\omega^{0,\beta}_{0,\alpha} + \omega^{0,\alpha}_{0,\beta})} & (-\omega^{\beta,0}_{\alpha,0}+\omega^{0,\beta}_{0,\alpha}) & (-\omega^{\alpha,0}_{\beta,0} + \omega^{0,\alpha}_{0,\beta})& (-\omega^{\alpha,0}_{\beta,0} - \omega^{\beta,0}_{\alpha,0}) & -\chi & 0 \\
{(\omega^{0,\alpha}_{0,\beta} + \omega^{0,\beta}_{0,\alpha})} & (-\omega^{\beta,0}_{\alpha,0}+\omega^{0,\beta}_{0,\alpha}) & (-\omega^{\alpha,0}_{\beta,0}+\omega^{0,\alpha}_{0,\beta}) & (-\omega^{\beta,0}_{\alpha,0} - \omega^{\alpha,0}_{\beta,0} ) & 0 & -\chi
\end{bmatrix}
\end{align}
where the terms in the boxes explicitly contains the counting fields as,
\begin{align}
    \omega_{\alpha,\beta} &= \omega^{\Leftarrow}_{\alpha,\beta}e^{\zeta} + \omega^{\Rightarrow}_{\alpha,\beta}e^{-\zeta} + \omega^{\rightleftarrows}_{\alpha,\beta} + \omega^{\leftrightarrows}_{\alpha,\beta} \\
    \omega_{\beta,\alpha} &= \omega^{\Leftarrow}_{\beta,\alpha}e^{\zeta} + \omega^{\Rightarrow}_{\beta,\alpha}e^{-\zeta} + \omega^{\rightleftarrows}_{\beta,\alpha} + \omega^{\leftrightarrows}_{\beta,\alpha} \\
    \omega_{0,2} &= \omega^{\Leftarrow}_{0,2} + \omega^{\Rightarrow}_{0,2} + \omega^{\leftrightarrow}_{0,2}e^{\zeta} \\
    \omega_{2,0} &= \omega^{\Leftarrow}_{2,0} + \omega^{\Rightarrow}_{2,0} + \omega^{\leftrightarrow}_{2,0}e^{-\zeta}.
\end{align}
The geometric flux obtained by using the left and right eigenvectors of Eq.(\ref{6x6 rate-exc}) is shown in Fig.(\ref{fig:j1g tl approx tr}e). This flux is solely due to the exchange cotunneling processes and is found to be negative and ten orders smaller than the flux due to sequential process, Fig.(\ref{fig:j1g tl approx tr}c) for the nonsecular case. For the secular case as seen in Fig.(\ref{fig:j1g tl approx tr}f), the exchange cotunneling flux is positive, comparable in magnitude and similar in shape to the total flux, Fig.(\ref{fig:j1g tl approx tr}b). Thus, we can say that for the secular case, the exchange cotunneling processes also influence the total flux whereas for the nonsecular case, the exchange cotunneling process do not influence the total geometric flux.

We next track the double charge-separated cotunneling processes. The twisted generator in this case is $\breve{\cal L}_{csc}$ given by
\begin{align}\label{6x6 rate double electron tracking}
\breve{\cal L}_{csc}=
\begin{bmatrix}
-\sum_{k\not=0}\omega_{0,k} & \omega_{\alpha,0} & \omega_{\beta,0} & \descbox{\omega_{2,0}} & {(\omega^{\beta,0}_{\alpha,0}+\omega^{\alpha,0}_{\beta,0})}& {(\omega^{\alpha,0}_{\beta,0}+\omega^{\beta,0}_{\alpha,0})} \\
{\omega_{0,\alpha}} & {-\sum_{k\not=\alpha} \omega_{\alpha,k}} &{\omega_{\beta,\alpha}} & {\omega_{2,\alpha}} & {(-\omega^{\alpha,0}_{\beta,0} + \omega^{0,\alpha}_{0,\beta})} & (-\omega^{\alpha,0}_{\beta,0}-\omega^{0,\alpha}_{0,\beta}) \\
{\omega_{0,\beta}} &  \omega_{\alpha,\beta} & -\sum_{k\not=\beta} \omega_{\beta,k} & \omega_{2,\beta} & (-\omega^{\beta,0}_{\alpha,0} + \omega^{0,\beta}_{0,\alpha} ) & (-\omega^{\beta,0}_{\alpha,0} + \omega^{0,\beta}_{0,\alpha}) \\
\descbox{\omega_{0,2}} &  \omega_{\alpha,2} & \omega_{\beta,2} & -\sum_{k\not=2} \omega_{2,k} & (- \omega^{0,\alpha}_{0,\beta} -\omega^{0,\beta}_{0,\alpha}) &( -\omega^{0,\alpha}_{0,\beta} - \omega^{0,\beta}_{0,\alpha}) \\
{(\omega^{0,\beta}_{0,\alpha} + \omega^{0,\alpha}_{0,\beta})} & (-\omega^{\beta,0}_{\alpha,0}+\omega^{0,\beta}_{0,\alpha}) & (-\omega^{\alpha,0}_{\beta,0} + \omega^{0,\alpha}_{0,\beta})& (-\omega^{\alpha,0}_{\beta,0} - \omega^{\beta,0}_{\alpha,0}) & -\chi & 0 \\
{(\omega^{0,\alpha}_{0,\beta} + \omega^{0,\beta}_{0,\alpha})} & (-\omega^{\beta,0}_{\alpha,0}+\omega^{0,\beta}_{0,\alpha}) & (-\omega^{\alpha,0}_{\beta,0}+\omega^{0,\alpha}_{0,\beta}) & (-\omega^{\beta,0}_{\alpha,0} - \omega^{\alpha,0}_{\beta,0} ) & 0 & -\chi
\end{bmatrix}
\end{align}
with the counting fields present as
\begin{align}
    \omega_{0,2} &= \omega^{\Leftarrow}_{0,2}e^{2\zeta} + \omega^{\Rightarrow}_{0,2} + \omega^{\leftrightarrow}_{0,2} \\
    \omega_{2,0} &= \omega^{\Leftarrow}_{2,0} + \omega^{\Rightarrow}_{2,0}e^{-2\zeta} + \omega^{\leftrightarrow}_{2,0}
\end{align}
\end{widetext}
The geometric flux obtained by using the left and right eigenvectors of Eq.(\ref{6x6 rate double electron tracking}) is shown in Fig.(\ref{fig:j1g tl approx tr}g) for the nonsecular case and the secular case, Fig.(\ref{fig:j1g tl approx tr}h). This flux has contribution only from the double charge-separated processes. For the nonsecular case this flux is found to be negative and ten orders smaller than the flux obtained from the sequential process, Fig.(\ref{fig:j1g tl approx tr}c). Therefore the double charge-transfer cotunneling processes have minimal contribution to the total geometric flux, Fig.(\ref{fig:j1g tl approx tr}a). For the secular case, Fig.(\ref{fig:j1g tl approx tr}h), the flux is positive, slightly smaller in magnitude but similar in shape to the total geometric flux, Fig.(\ref{fig:j1g tl approx tr}b). It can be concluded that the total geometric flux, Fig.(\ref{fig:j1g tl approx tr}b) is dominated by the exchange cotunneling processes (as was seen in Fig.(\ref{fig:j1g tl approx tr}f)) with contribution from double charge-separated cotunneling processes as well.  
Thus, upon comparing the four different fluxes in Fig.(\ref{fig:j1g tl approx tr}c, \ref{fig:j1g tl approx tr}d, \ref{fig:j1g tl approx tr}e, \ref{fig:j1g tl approx tr}f, \ref{fig:j1g tl approx tr}g and \ref{fig:j1g tl approx tr}h) using the four different tracker Liouvillians,[EQ...] we can conclude the following. For the nonsecular case, sequential processes dominate the total geometric flux and the cotunneling processes (be it exchange or double charge-separated) have minimal role to play. This is physically acceptable because the sequential processes are more in number than the cotunneling processes (the coherences contain only the sequential terms). For the secular case, the number of sequential processes reduce (since the coherences vanish) and the contribution from the cotunneling processes increase. Therefore the cotunneling contribution to the total geometric flux is increased. In contrast to the linear dependence in interacting quantum dots\cite{yuge2012geometrical}, the dependence of the four different types of fluxes on the interaction energy is found to be nonlinear.

The above discussion is valid when the coupling of the spinorbitals with the left and right reservoirs are unequal but comparable. It has been previously shown that the cotunneling contributions to the dynamic flux in spinless double quantum dot junctions can be increased by constructing asymmetrical system-reservoir couplings strength, i.e $t_L\gg t_R$ or $t_L \ll t_R$ \cite{carmi2012enhanced}. We investigate the geometric flux under these two conditions in the similar fashion to what was done for the $t_L\approx t_R$ case above.
\\
\begin{figure}
\centering
    \includegraphics[width = 8.5cm]{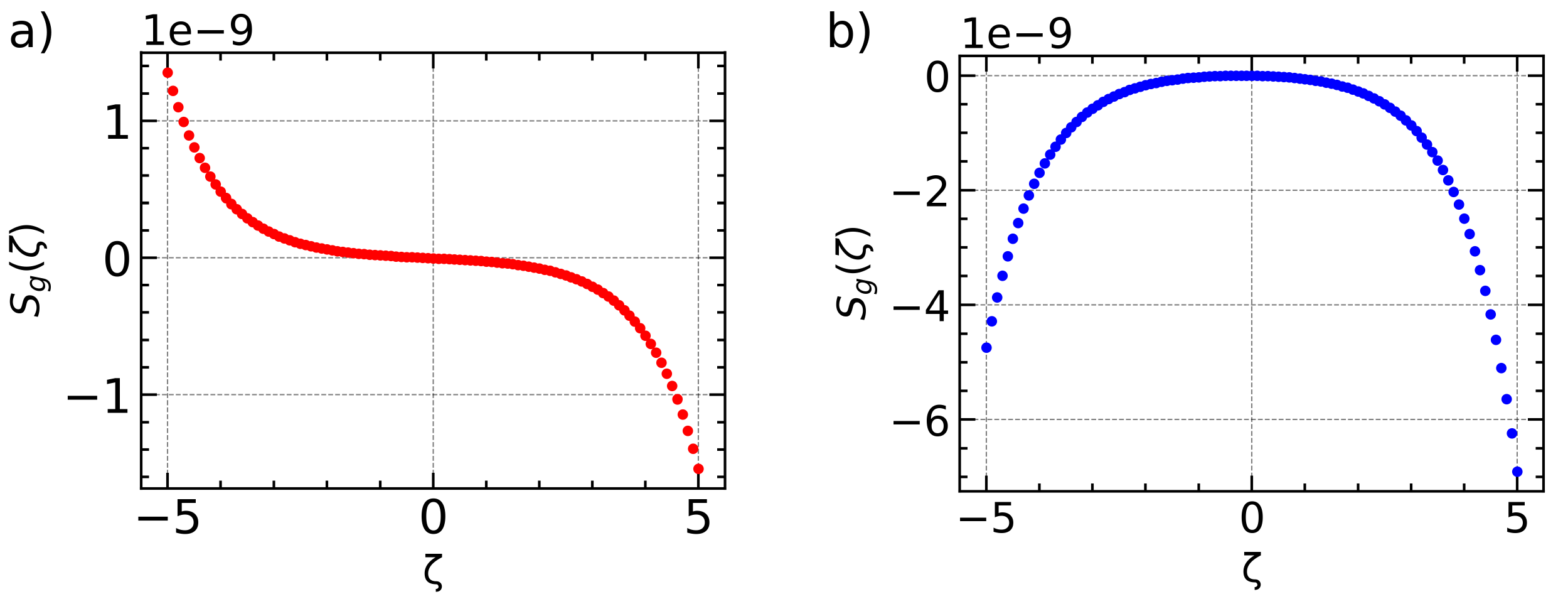}
    \caption{Behavior of the geometric cumulant generating function in the presence of cotunneling, $S_{g}(\zeta)$ obtained using Eq.(\ref{6x6 rate}) and Eq. (\ref{eq:geometric part = line intergal time}) for both (a) non-secular and (b) secular case vs the auxiliary counting field $\zeta$ for $t_L\gg t_R$ with the following parameters $[\Gamma = 2(\Gamma_{L} + \Gamma_{R})]$ : $K_{B}T = \hbar\Gamma, E_{\alpha} = 2 meV, E_{\beta} = E_{\alpha} - 0.7\hbar\Gamma, U = 1.5meV, \mu^{\circ}_{L} = 4meV, eV_{bias} = 3\hbar\Gamma, m = 0.01$ and $\phi = \frac{\pi}{3}$.}
    \label{fig:GP tl great tr}
\end{figure}
\begin{figure}
    \centering
    \includegraphics[width =8.5cm]{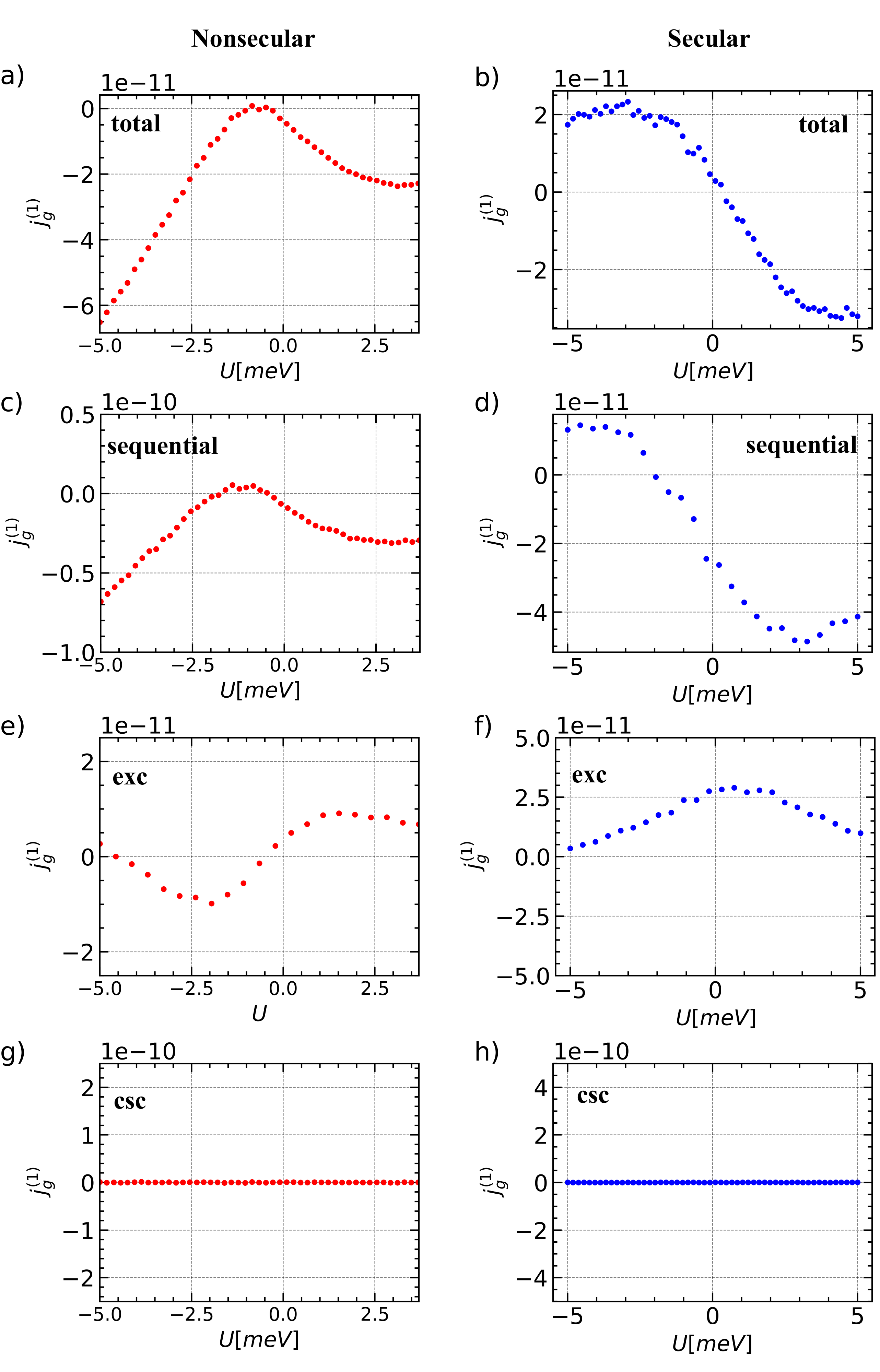}
    \caption{(Color Online) Behavior of the total geometric flux, $j^{(1)}_g$ (Eq.(\ref{eq:jkg}) with $k=1$ and $t_{L}\gg t_{R}$) as a function of the interaction potential $U$ obtained using Eqs.(\ref{eq:5}-\ref{eq-hybrid-rates-}) in the superoperator, Eq.(\ref{6x6 rate}) to calculate Eq. (\ref{eq:geometric part = line intergal time})   (a) in the presence  and (b) absence of coherence. 
    Plots in red (left side) and blue (right side) correspond to the non secular and secular case respectively.
    Contribution to the geometric flux from only the sequential electron transfer process (from Eq.(\ref{6x6 rate-seq})), in presence (c) and absence (d) of coherences. Contribution only from the exchange cotunneling processes (from Eq. (\ref{6x6 rate-exc})) in (e) presence (e) and (f) absence of coherences. Contribution from only the double charge-separated processes (from Eq.(\ref{6x6 rate double electron tracking})) are shown in (g) with coherences and (h) without coherences.}
    \label{fig:j1g tl great tr}
\end{figure}
We first consider the case when $t_L\gg t_R$. In this case, the geometric cumulant generating function for the nonsecular and secular case is shown in Fig.(\ref{fig:GP tl great tr}a) and Fig.(\ref{fig:GP tl great tr}b) respectively. 
For both cases, the magnitude of geometric cumulant generating functions is several orders smaller than the case when $t_L\approx t_R$.  Hence we expect lower magnitude of the flux in this case.  
In Fig.(\ref{fig:j1g tl great tr}a and b) we show the variation of total geometric flux as a function of interaction energy $U$ when $t_L\gg t_R$ for the nonsecular and secular case respectively.
For the nonsecular case (plotted red), Fig.(\ref{fig:j1g tl great tr}a) the total geometric flux increases almost linearly in the region of $U > 0$ reaching a maxima, thereafter decreases as the repulsion energy increases. In Fig.(\ref{fig:j1g tl great tr}c and d) we evaluate the geometric flux for the sequential processes alone. In Fig.(\ref{fig:j1g tl great tr}e and f) we evaluate the geometric flux for the exchange cotunneling processes. As can be seen from Fig.(\ref{fig:j1g tl great tr}c,d,e and f) both the sequential and exchange cotunneling processes contribute to the total geometric flux. The contribution from double charge-separated cotunneling processes is absent when $t_L \gg t_R$ as seen in Fig.(\ref{fig:j1g tl great tr}g and h).
This result is acceptable. The coupling in the right junction is very weak compared to the left junction and hence the probability that the two electrons simultaneously get transported to the right reservoir is low. Since the tracking involves the system and right junction, the contribution to the flux from such processes is negligible and we see a flat line. We safely conclude that, maintaining $t_L\gg t_R$ doesn't increase the cotunneling contributions to the geometric flux contrary to what is known for dynamic flux \cite{carmi2012enhanced}.
\begin{figure}
   \centering
    \includegraphics[width =8.5cm]{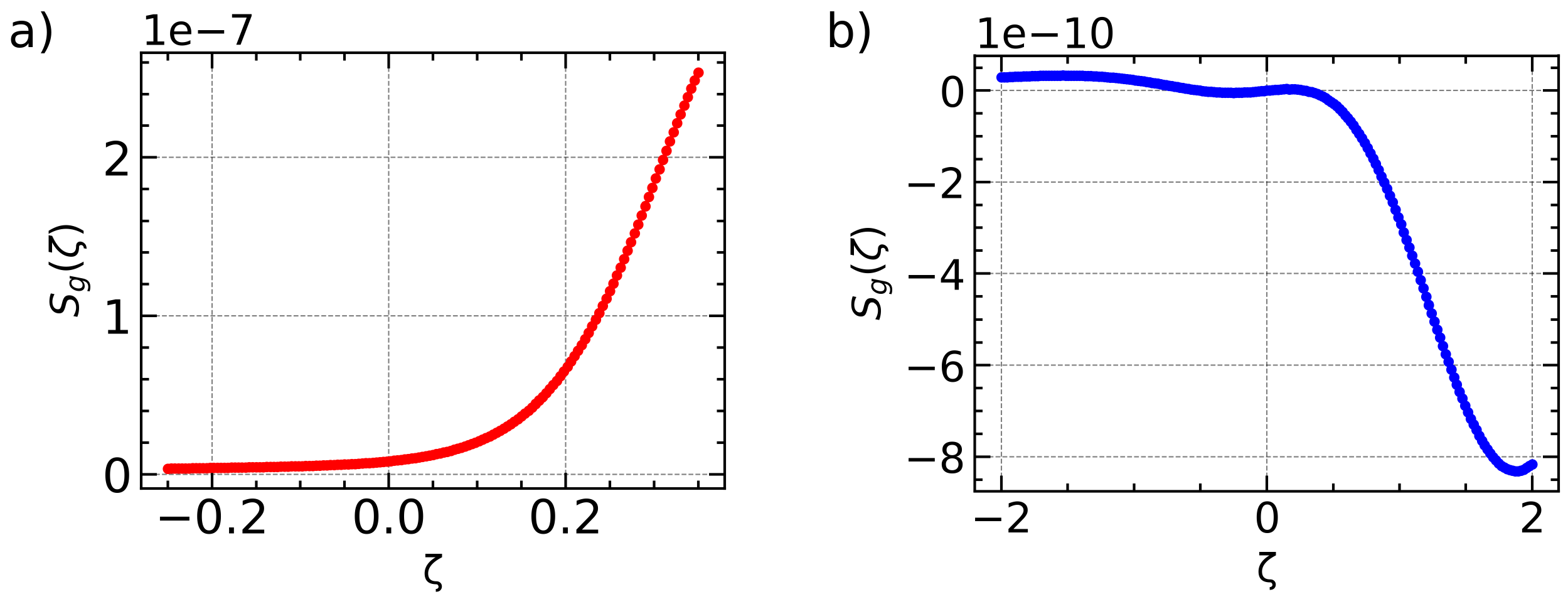}
   \caption{Behavior of the geometric cumulant generating function in the presence of cotunneling, $S_{g}(\zeta)$ obtained using Eq.(\ref{6x6 rate}) and Eq.(\ref{eq:geometric part = line intergal time})for both (a) non-secular and (b) secular case vs the auxiliary counting field $\zeta$ for $t_L\ll t_R$ with the following parameters $[\Gamma = 2(\Gamma_{L} + \Gamma_{R})]$ : $K_{B}T = \hbar\Gamma, E_{\alpha} = 4 meV, E_{\beta} = E_{\alpha} - 0.1\hbar\Gamma, U = 1.5meV, \mu^{\circ}_{L} = 5meV, eV_{bias} = 3\hbar\Gamma, m = 0.01$ and $\phi = \frac{\pi}{3}$.}
    \label{fig:GP tl less tr}
\end{figure}
\begin{figure}
    \centering
    \includegraphics[width =8.5cm]{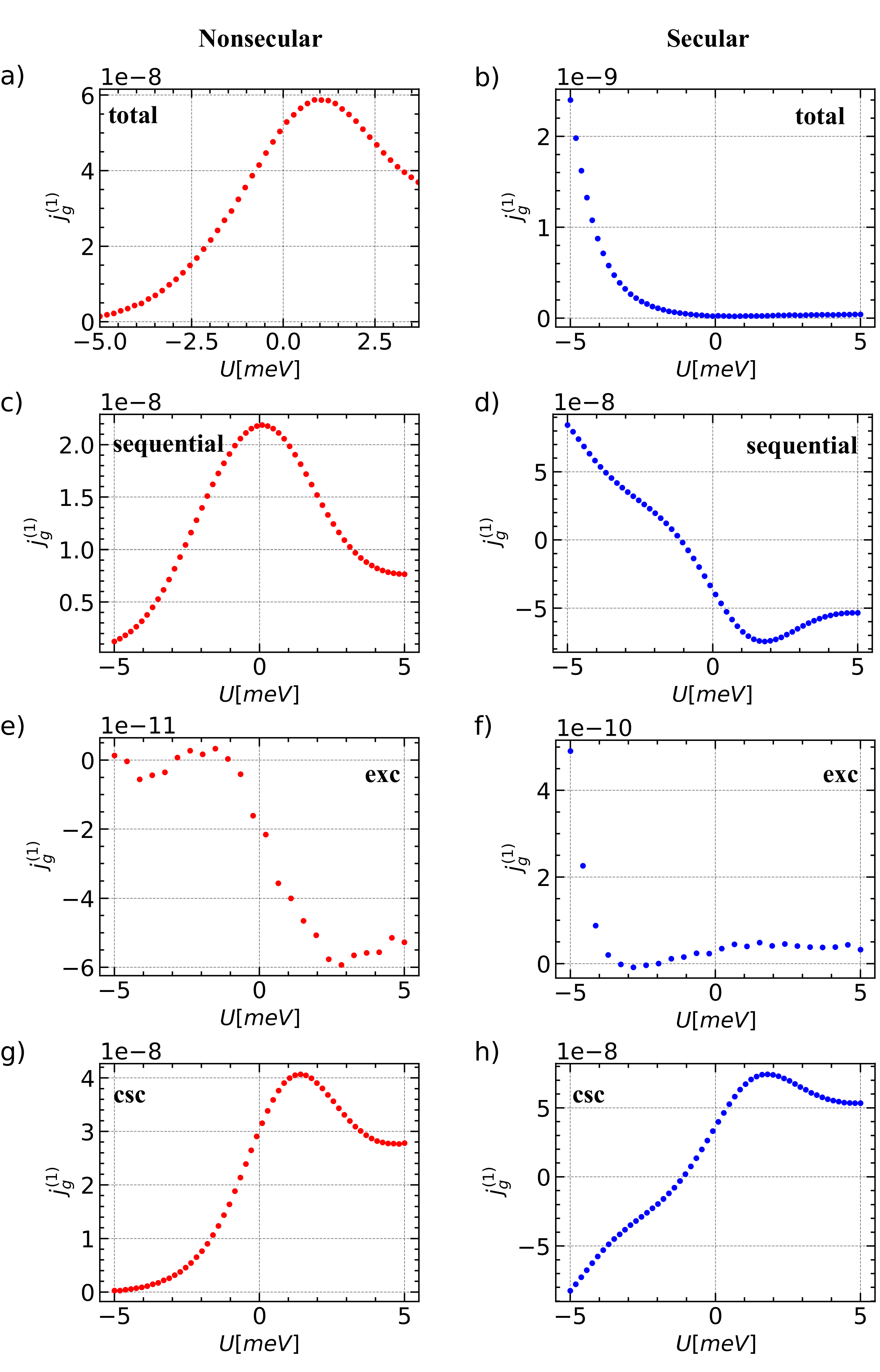}
    \caption{(Color Online) Behavior of the total geometric flux, $j^{(1)}_g$  (from Eq. (\ref{eq:jkg}) with $k=1$ and $t_{L}\ll t_{R}$) as a function of the interaction potential $U$ obtained using Eqs. (\ref{eq:5}-\ref{eq-hybrid-rates-}) in the superoperator, Eq.(\ref{6x6 rate}) to calculate Eq. (\ref{eq:geometric part = line intergal time}) (a) in the presence  and (b) absence of coherence. 
    Plots in red (left side) and blue (right side) correspond to the non secular and secular case respectively. 
    Contribution to the geometric flux from only the sequential electron transfer process (from Eq.(\ref{6x6 rate-seq})), in presence (c) and absence (d) of coherences. Contribution only from the exchange cotunneling processes (from Eq. (\ref{6x6 rate-exc})) in (e) presence (e) and (f) absence of coherences. Contribution only from the double charge-separated processes (from Eq.(\ref{6x6 rate double electron tracking})) are shown in (g) with coherences and (h) without coherences.}
    \label{fig:j1g tl less tr}
\end{figure}

We now explore the other extreme condition, $t_L\ll t_R$, which is also known to increase the cotunneling contributions to the dynamic flux \cite{carmi2012enhanced}. Following the earlier protocol, the cumulant generating functions are shown in Fig.(\ref{fig:GP tl less tr}a and \ref{fig:GP tl less tr}b) for both non-secular and secular case respectively. The magnitude of the geometric cumulant generating function in the presence of coherences is much higher than the secular case. In this limit, the coherences enhance the overall geometric flux as evident in Fig.(\ref{fig:j1g tl less tr}a and b). The contributions from the sequential, exchange cotunneling and double charge-separated cotunneling processes to the total geometric flux is shown in Fig.(\ref{fig:j1g tl less tr}c-h). For the nonsecular case, the double charge-separated cotunneling processes have a twofold larger contribution (Fig.(\ref{fig:j1g tl less tr}g)) than the sequential processes (Fig.(\ref{fig:j1g tl less tr}c)) to the total geometric flux (Fig.(\ref{fig:j1g tl less tr}a)). The exchange cotunneling processes do not contribute to the total flux since it is several orders lower in magnitude, Fig.(\ref{fig:j1g tl less tr}e) than the other two processes. From the shape and magnitude of the curves in Fig.(\ref{fig:j1g tl less tr}d) and Fig.(\ref{fig:j1g tl less tr}h), we can infer that the sequential and double charge-separated cotunneling processes nearly cancel each other out for the secular case. The major contributing factor to the total flux are the exchange cotunneling processes, Fig.(\ref{fig:j1g tl less tr}f) which is of the order of $10^{-10}$.
Thus, upon comparing the fluxes for the condition $t_{L} \ll t_{R}$ for four different tracking Liouvillian, Fig.(\ref{fig:j1g tl less tr}c, \ref{fig:j1g tl less tr}d, \ref{fig:j1g tl less tr}e, \ref{fig:j1g tl less tr}f, \ref{fig:j1g tl less tr}g and \ref{fig:j1g tl less tr}h), it can be concluded that, for non-secular case, the double charge-separated cotunneling and sequential cotunneling, processes dominate the net geometric flux in the entire range of $U$. Whereas, for secular case, only exchange cotunneling processes contribute to the net geometric flux.
\begin{figure}
    \centering 
    \includegraphics[width =8.5cm]{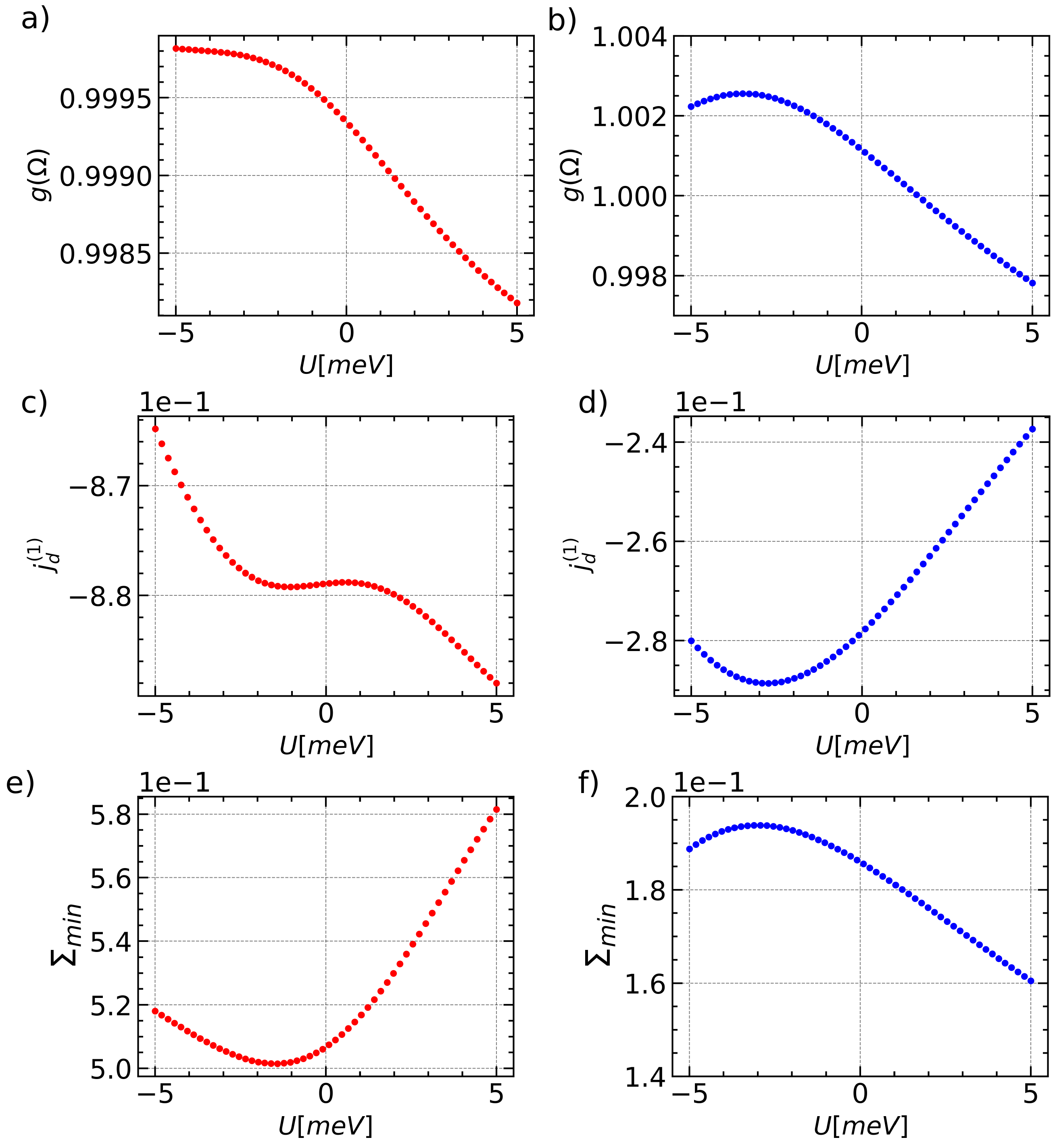}
    \caption{Behavior of the geometric geometric correction factor (a) nonsecular (b) secular when $t_L\approx t_R$. Minimal entropy production as a function of $U$ for (c) non-secular (d) secular case.}
    \label{fig:entropy}
\end{figure}
\section{Estimation of Entropy}
For an nondriven case, $S_g(\zeta)=0 $ (when $\Omega =0$ or $\phi = 0$) and the overall statistics of the transport across the spintronic junction is governed by the dynamic component. As such, a thermodynamic uncertainty relationship (TUR) obtainable from a steadystate fluctuation theorem holds, given by $F{\cal A}\ge 2k_B $ \cite{pietzonka2017finite,saryal2019thermodynamic} with  $F$ being the  Fano factor (ratio between the total second and first cumulant) ${\cal A}$ is the thermodynamic affinity  of the system and is directly proportional to the steadystate entropy of the junction.
Such a TUR has been shown not to hold in the presence of geometric effects \cite{PhysRevE.96.052129} and 
has been modified by including a geometric correction factor. The modified relationship is of the form, 
$
 (j^{(2)}_g+j^{(2)}_d)\Sigma/(g(\Omega)(j^{(1)}_d+j^{(2)}_g))^2\ge 2k_B 
$\cite{lu2022geometric}. $\Sigma$ is the total entropy production rate and has inseparable contributions from both geometric and dynamic components. 
$g(\Omega)$ is the driving dependent geometric correction factor containing both the dynamic and geometric flux in a nonlinear fashion given by,
\begin{align}
\label{eq-geo-tur}
g(\Omega)&=\frac{1}{(1+j^{(1)}_g/j^{(1)}_d)^2}.
\end{align}
The $k$-th dynamic cumulant can be obtained using,
\begin{align}
 j^{(k)}_d&=
 \displaystyle\frac{1}{t_p}\int_0^{t_p}\frac{d^{k}}{d\zeta^k} \lambda_o(\zeta,t)|_{\zeta=0}(t) dt.
\end{align}
Note that, in the presence of geometricities, evaluation of entropies with contribution from both dynamic and geometric components is not at all straightforward \cite{PhysRevE.106.064119} due to the production of excess entropies\cite{yuge2013geometrical}. However, using this modified TUR, we can estimate the minimum entropy production rate as follows,
\begin{align}
\label{eq-entropy}
 \Sigma_{min}= 2k_B\frac{(j^{(1)}_d+j^{(1)}_g)^2}{j^{(2)}_d+j^{(2)}_g}g(\Omega)
\end{align}
We numerically evaluate Eq.(\ref{eq-geo-tur}) and plot $g(\Omega)$ as a function of $U$ for nonsecular (secular) in Fig.(\ref{fig:entropy}a(b)). We also also evaluate Eq.(\ref{eq-entropy}) and plot it as a function of $U$ for both the nonsecular and secular cases in Fig.(\ref{fig:entropy}c) and Fig.(\ref{fig:entropy}d) respectively. 
For the non secular case, as seen in Fig.(\ref{fig:entropy}a), $g(\Omega) < 1$ in the entire region of $U$ and decreases gradually as  $U$ changes from being attractive to repulsive. Whereas for the secular case, as seen in Fig.(\ref{fig:entropy}b), $g(\Omega) > 1$ with a maximum value at $U \approx -3.25 meV$ and then decreases linearly. From $U \approx 1.65 meV$, $g(\Omega) < 1$. In Fig.(\ref{fig:entropy}a) $g(\Omega)$ is always less than unity since both the dynamic (Fig.(\ref{fig:entropy}c)) and geometric flux are negative. This makes the denominator of the lhs term of Eq.(\ref{eq-geo-tur}) larger than unity. The larger than unity behaviour of $g(\Omega)$ in Fig.(\ref{fig:entropy}b) is because of the fact that the dynamic flux is negative whereas the geometric flux is positive in the region $-5< U \le 1.65$ (see Fig.(\ref{fig:j1g tl approx tr})b for the geometric flux). This makes the denominator in the lhs of Eq.(\ref{eq-geo-tur}) less than unity. From $U=1.65meV$ onwards the geometric flux is negative resulting in $g(\Omega)<1$. Thus the coherences change the nature of the geometric correction factor. 

The minimum entropy production for the non secular case is shown in Fig.(\ref{fig:entropy}c). It shows a decreasing nature in the region of $U < 0$ until it hits a minima at $U \approx 1.5$. It then starts decreasing almost linearly for $U > 0$. Whereas, for the secular case, the minimum entropy production rate increases until it hits a maximum at $U \approx 2.6 meV$ and then decreases in a linear fashion as $U$ changes from being attractive to repulsive. Here too, the coherences completely change the nature of the minimal entropy production rate.  
\section{Conclusion}

We systematically investigated the geometric statistics of a spintronic junction using principles from full counting statistical framework by adiabatically modulating the chemical potential of the reservoirs in the spin resolved basis. We take into account both  sequential as well as different types of cotunneling processes during the nonequilibrium spin transfer across the junction.  By separately tracking the total, sequential, exchange cotunneling and double charge-separated cotunneling processes we numerically evaluated the geometric flux as a function of both attractive and repulsive spin-spin interaction energy. We investigated this both in the presence and absence of coherences by constructing analytical superoperators describing the overall spin transport process. The nature of the total geometric flux as a function of the interaction energy is very different in presence and absence of coherences. In contrast to linear dependence of the geometric flux on the Coulomb electron-electron (spinless) repulsion energy in double quantum dots, we find that the dependence of the geometric flux on the total spin-spin interaction energy is nonlinear. We identified relevant conditions when the cotunneling processes contribute to the total geometric flux. We found that when the strength of the left and the right system-reservoir couplings are comparable, cotunneling effects influence the total flux only in the absence of coherences. When the left reservoir coupling is stronger than the right reservoir coupling, cotunneling of spins have no effect on the overall geometric flux both in the presence and absence of coherence. When the right reservoir-system coupling is larger than the left system-reservoir coupling, the sequential and the double charge-separated cotunneling processes cancel each other out. In this case the cotunneling processes which involve simultaneous exchanges with both left and right reservoirs contribute to the overall geometric flux. Upon exploring a recently proposed geometric thermodynamic uncertainty relationship we found that the coherences completely alter the nature of the dependence of the interaction energy on the geometric correction factor and the minimal entropy production rate. 

\section*{Acknowledgments}
   MS and JA appreciate the support from the Department of Chemistry, Gauhati University. MJS thanks the Science and Engineering Research Board for the fellowship from the grant with file number SERB/SRG/2021/001088. HPG acknowledges the support from the University Grants Commission, New Delhi for the startup research grant, UGC(BSR), Grant No.  F.30-585/2021(BSR).  

\section*{Appendix: Analytical Expressions for the sequential and cotunneling rates}

The expressions of the sequential tunneling process are:
\begin{align}
    \omega^{\rightarrow}_{0,\alpha} &= \Gamma_L F_{FD}(E_{\alpha} - \mu_L) \\
    \omega^{\rightarrow}_{0,\beta} &= \Gamma_L F_{FD}(E_{\beta} - \mu_L) \\
    \omega^{\rightarrow}_{\alpha,0} &= \Gamma_R F_{FD}(\mu_R - E_{\alpha}) \\
    \omega^{\rightarrow}_{\beta,0} &= \Gamma_R F_{FD}(\mu_R - E_{\beta}) \\ 
    \omega^{\rightarrow}_{\alpha,2} &= \Gamma_L F_{FD}(E_{\alpha} + U - \mu_L) \\
    \omega^{\rightarrow}_{\beta,2} &= \Gamma_L F_{FD}(E_{\beta} + U - \mu_L) \\
    \omega^{\rightarrow}_{2,\beta} &= \Gamma_R F_{FD}(\mu_R - E_{\alpha} - U) \\
    \omega^{\rightarrow}_{2,\alpha} &= \Gamma_R F_{FD}(\mu_R - E_{\beta} - U)
\end{align}
The conjugate processes towards the left can be obtained using the similar expression by taking the transformation $\mu_{_L} \leftrightarrow \mu_{_R}$ and $\Gamma_{L} \leftrightarrow \Gamma_{R}$.
\\
\begin{widetext}
The expression of the exchange cotunneling process are:
\begin{align} \label{eq:exchange cotunneling process1}
    \omega_{\alpha,\beta}^{\Rightarrow} &= \frac{2\pi\nu^{2}}{\hbar}{\int}F_{FD}(\varepsilon - \mu_{L})[1-F_{FD}(\varepsilon -(E_{\beta}-E_{\alpha}+\mu_{R})]{\bigg|{\frac{t_{R}t_{L}}{\varepsilon - E_{\beta}}} - {\frac{t_{R}t_{L}}{\varepsilon - (E_{\beta} + U)}}\bigg|}^2{d\varepsilon} \\
    \omega_{\alpha,\beta}^{\Leftarrow} &= \frac{2\pi\nu^{2}}{\hbar}{\int}F_{FD}(\varepsilon - \mu_{R})[1-F_{FD}(\varepsilon -(E_{\beta} - E_{\alpha} + \mu_{L})]\bigg|{\frac{t_{R}t_{L}}{\varepsilon - E_{\beta}}} - {\frac{t_{R}t_{L}}{\varepsilon - (E_{\beta} + U)}}\bigg|^2{d\varepsilon} \\
    \omega_{\alpha,\beta}^{\rightleftarrows} &= \frac{2\pi\nu^{2}}{\hbar}{\int}F_{FD}(\varepsilon - \mu_{R})[1-F_{FD}(\varepsilon -(E_{\beta} - E_{\alpha}+ \mu_{R})]\bigg|{\frac{t_{R}t_{R}}{\varepsilon - E_{\beta}}} - {\frac{t_{R}t_{R}}{\varepsilon - (E_{\beta} + U)}}\bigg|^2{d\varepsilon} \\
    \label{eq:exchange cotunneling process4}
    \omega_{\alpha,\beta}^{\leftrightarrows} &= \frac{2\pi\nu^{2}}{\hbar}{\int}F_{FD}(\varepsilon - \mu_{L})[1-F_{FD}(\varepsilon -(E_{\beta} - E_{\alpha}+ \mu_{L})]\bigg|{\frac{t_{L}t_{L}}{\varepsilon - E_{\beta}}} - {\frac{t_{L}t_{L}}{\varepsilon - (E_{\beta} + U)}}\bigg|^2{d\varepsilon}
\end{align}
To obtain the rates of $\omega_{\beta\alpha}^{\Rightarrow},\omega_{\beta\alpha}^{\Leftarrow},\omega_{\beta\alpha}^{\rightleftarrows},\omega_{\alpha\beta}^{\leftrightarrows}$, the integrals from the Eq.(\ref{eq:exchange cotunneling process1}) to (\ref{eq:exchange cotunneling process4}) can be used along with the transformation $E_{\alpha} \leftrightarrow E_{\beta}$. 
\\
The rate expression for double charge-separated cotunneling processes are 
\begin{align}
    \omega_{0,2}^{\Rightarrow} &= \frac{2\pi\nu^{2}}{\hbar}{\int}\frac{1}{2}F_{FD}(\varepsilon-\mu_{L})F_{FD}[(-\varepsilon+ E_{\alpha}+E_{\beta}+U)-\mu_{L}]\bigg|{\frac{t_{L}t_{L}}{\varepsilon-E_{\alpha}}} - {\frac{t_{L}t_{L}}{\varepsilon - E_{\alpha} - U}} - {\frac{t_{L}t_{L}}{\varepsilon-E_{\beta}}}+ {\frac{t_{L}t_{L}}{\varepsilon - E_{\beta} - U}}\bigg|^2{d\varepsilon}, \\
    \omega_{0,2}^{\Leftarrow} &= \frac{2\pi\nu^{2}}{\hbar}{\int}\frac{1}{2}F_{FD}(\varepsilon-\mu_{R})F_{FD}[(-\varepsilon+ E_{\alpha}+E_{\beta}+U)-\mu_{R}]\bigg|{\frac{t_{R}t_{R}}{\varepsilon-E_{\alpha}}} - {\frac{t_{R}t_{R}}{\varepsilon - E_{\alpha} -U}} - {\frac{t_{R}t_{R}}{\varepsilon-E_{\beta}}} + {\frac{t_{R}t_{R}}{\varepsilon - E_{\beta} - U}}\bigg|^2{d\varepsilon},\\
    \omega_{0,2}^{\leftrightarrow} &= \frac{2\pi\nu^{2}}{\hbar}{\int}F_{FD}(\varepsilon-\mu_{L})F_{FD}[(-\varepsilon+ E_{\alpha}+E_{\beta}+U)-\mu_{R}]\bigg|{\frac{t_{L}t_{R}}{\varepsilon-E_{\alpha}}} - {\frac{t_{L}t_{R}}{\varepsilon - E_{\alpha}+U}} - {\frac{t_{L}t_{R}}{\varepsilon-E_{\beta}}} + {\frac{t_{L}t_{R}}{\varepsilon - E_{\beta} + U}}\bigg|^2{d\varepsilon} \\
    \omega_{2,0}^{\Rightarrow} &= \frac{2\pi\nu^{2}}{\hbar}{\int}\frac{1}{2}F_{FD}(\mu_{L}-\varepsilon)F_{FD}[\mu_L-(-\varepsilon + E_{\alpha} + E_{\beta} + U)]\bigg|{\frac{t_{R}t_{R}}{\varepsilon-E_{\alpha}}} - {\frac{t_{R}t_{R}}{\varepsilon - E_{\alpha}-U}} - {\frac{t_{R}t_{R}}{\varepsilon-E_{\beta}}} + {\frac{t_{R}t_{R}}{\varepsilon - E_{\beta} - U}}\bigg|^2{d\varepsilon}, \\
    \omega_{2,0}^{\Leftarrow} &= \frac{\pi\nu^{2}}{\hbar}{\int}\frac{1}{2}F_{FD}(\mu_{R}-\varepsilon)F_{FD}[\mu_R-(-\varepsilon + E_{\alpha} + E_{\beta} + U)]\bigg|{\frac{t_{L}t_{L}}{\varepsilon-E_{\alpha}}} - {\frac{t_{L}t_{L}}{\varepsilon - E_{\alpha}-U}} - {\frac{t_{L}t_{L}}{\varepsilon-E_{\beta}}} + {\frac{t_{L}t_{L}}{\varepsilon - E_{\beta} - U)}}\bigg|^2{d\varepsilon},\\
    \omega_{2,0}^{\leftrightarrow} &= \frac{2\pi\nu^{2}}{\hbar}{\int}F_{FD}(\mu_L - \varepsilon)F_{FD}[\mu_R-(-\varepsilon + E_{\alpha} + E_{\beta} + U)]\bigg|{\frac{t_{L}t_{R}}{\varepsilon-E_{\alpha}}} - {\frac{t_{L}t_{R}}{\varepsilon - E_{\alpha}-U}} - {\frac{t_{L}t_{R}}{\varepsilon-E_{\beta}}} + {\frac{t_{L}t_{R}}{\varepsilon - E_{\beta}- U}}\bigg|^2{d\varepsilon}
\end{align}

The rate expressions double charge-separated cotunneling processes are
\begin{align}
    \omega_{0,2}^{\Rightarrow} &= \frac{2\pi\nu^{2}}{\hbar}{\int}\frac{1}{2}F_{FD}(\varepsilon-\mu_{L})F_{FD}[(-\varepsilon+ E_{\alpha}+E_{\beta}+U)-\mu_{L}]\bigg|{\frac{t_{L}t_{L}}{\varepsilon-E_{\alpha}}} - {\frac{t_{L}t_{L}}{\varepsilon - E_{\alpha} - U}} - {\frac{t_{L}t_{L}}{\varepsilon-E_{\beta}}}+ {\frac{t_{L}t_{L}}{\varepsilon - E_{\beta} - U}}\bigg|^2{d\varepsilon}, \\
    \omega_{0,2}^{\Leftarrow} &= \frac{2\pi\nu^{2}}{\hbar}{\int}\frac{1}{2}F_{FD}(\varepsilon-\mu_{R})F_{FD}[(-\varepsilon+ E_{\alpha}+E_{\beta}+U)-\mu_{R}]\bigg|{\frac{t_{R}t_{R}}{\varepsilon-E_{\alpha}}} - {\frac{t_{R}t_{R}}{\varepsilon - E_{\alpha} -U}} - {\frac{t_{R}t_{R}}{\varepsilon-E_{\beta}}} + {\frac{t_{R}t_{R}}{\varepsilon - E_{\beta} - U}}\bigg|^2{d\varepsilon},
\end{align}
\begin{align}
    \omega_{0,2}^{\leftrightarrow} &= \frac{2\pi\nu^{2}}{\hbar}{\int}F_{FD}(\varepsilon-\mu_{L})F_{FD}[(-\varepsilon+ E_{\alpha}+E_{\beta}+U)-\mu_{R}]\bigg|{\frac{t_{L}t_{R}}{\varepsilon-E_{\alpha}}} - {\frac{t_{L}t_{R}}{\varepsilon - E_{\alpha}+U}} - {\frac{t_{L}t_{R}}{\varepsilon-E_{\beta}}} + {\frac{t_{L}t_{R}}{\varepsilon - E_{\beta} + U}}\bigg|^2{d\varepsilon} \nonumber\\
    &+ \frac{2\pi\nu^{2}}{\hbar}{\int}F_{FD}(\varepsilon-\mu_{R})F_{FD}[(-\varepsilon+ E_{\alpha}+E_{\beta}+U)-\mu_{L}]\bigg|{\frac{t_{L}t_{R}}{\varepsilon-E_{\alpha}}} - {\frac{t_{L}t_{R}}{\varepsilon - E_{\alpha}+U}} - {\frac{t_{L}t_{R}}{\varepsilon-E_{\beta}}} + {\frac{t_{L}t_{R}}{\varepsilon - E_{\beta} + U}}\bigg|^2{d\varepsilon}
\end{align}
\begin{align}
    \omega_{2,0}^{\Rightarrow} &= \frac{2\pi\nu^{2}}{\hbar}{\int}\frac{1}{2}F_{FD}(\mu_{L}-\varepsilon)F_{FD}[\mu_L-(-\varepsilon + E_{\alpha} + E_{\beta} + U)]\bigg|{\frac{t_{R}t_{R}}{\varepsilon-E_{\alpha}}} - {\frac{t_{R}t_{R}}{\varepsilon - E_{\alpha}-U}} - {\frac{t_{R}t_{R}}{\varepsilon-E_{\beta}}} + {\frac{t_{R}t_{R}}{\varepsilon - E_{\beta} - U}}\bigg|^2{d\varepsilon}, \\
    \omega_{2,0}^{\Leftarrow} &= \frac{\pi\nu^{2}}{\hbar}{\int}\frac{1}{2}F_{FD}(\mu_{R}-\varepsilon)F_{FD}[\mu_R-(-\varepsilon + E_{\alpha} + E_{\beta} + U)]\bigg|{\frac{t_{L}t_{L}}{\varepsilon-E_{\alpha}}} - {\frac{t_{L}t_{L}}{\varepsilon - E_{\alpha}-U}} - {\frac{t_{L}t_{L}}{\varepsilon-E_{\beta}}} + {\frac{t_{L}t_{L}}{\varepsilon - E_{\beta} - U)}}\bigg|^2{d\varepsilon}
\end{align}
\begin{align}
    \omega_{2,0}^{\leftrightarrow} &= \frac{2\pi\nu^{2}}{\hbar}{\int}F_{FD}(\mu_L - \varepsilon)F_{FD}[\mu_R-(-\varepsilon + E_{\alpha} + E_{\beta} + U)]\bigg|{\frac{t_{L}t_{R}}{\varepsilon-E_{\alpha}}} - {\frac{t_{L}t_{R}}{\varepsilon - E_{\alpha}-U}} - {\frac{t_{L}t_{R}}{\varepsilon-E_{\beta}}} + {\frac{t_{L}t_{R}}{\varepsilon - E_{\beta}- U}}\bigg|^2{d\varepsilon}\\ 
    \nonumber
    &+ \frac{2\pi\nu^{2}}{\hbar}{\int}F_{FD}(\mu_R - \varepsilon)F_{FD}[\mu_L-(-\varepsilon + E_{\alpha} + E_{\beta} + U)]\bigg|{\frac{t_{L}t_{R}}{\varepsilon-E_{\alpha}}} - {\frac{t_{L}t_{R}}{\varepsilon - E_{\alpha}-U}} - {\frac{t_{L}t_{R}}{\varepsilon-E_{\beta}}} + {\frac{t_{L}t_{R}}{\varepsilon - E_{\beta}- U}}\bigg|^2{d\varepsilon}
\end{align}

The rate expression for double charge-separated cotunneling processes are: 
\begin{align}
    \omega_{0,2}^{\Rightarrow} &= \frac{\pi\nu^{2}}{\hbar}{\int}F_{FD}(\varepsilon-\mu_{L})[1-F_{FD}(\varepsilon-(E_{\alpha}+E_{\beta}+U-\mu_{L}))]\bigg|{\frac{t_{L}t_{L}}{\varepsilon-E_{\alpha}}} - {\frac{t_{L}t_{L}}{\varepsilon - (E_{\alpha}+ U)}} - {\frac{t_{L}t_{L}}{\varepsilon-E_{\beta}}}+ {\frac{t_{L}t_{L}}{\varepsilon - (E_{\beta}+ U)}}\bigg|^2{d\varepsilon}, \\
    \omega_{0,2}^{\Leftarrow} &= \frac{\pi\nu^{2}}{\hbar}{\int}F_{FD}(\varepsilon-\mu_{R})[1-F_{FD}(\varepsilon-(E_{\alpha}+E_{\beta}+U-\mu_{R}))]\bigg|{\frac{t_{R}t_{R}}{\varepsilon-E_{\alpha}}} - {\frac{t_{R}t_{R}}{\varepsilon - (E_{\alpha}+U)}} - {\frac{t_{R}t_{R}}{\varepsilon-E_{\beta}}} + {\frac{t_{R}t_{R}}{\varepsilon - (E_{\beta}+ U)}}\bigg|^2{d\varepsilon},\\
    \omega_{0,2}^{\leftrightarrow} &= \frac{2\pi\nu^{2}}{\hbar}{\int}F_{FD}(\varepsilon-\mu_{L})[1-F_{FD}(\varepsilon-(E_{\alpha}+E_{\beta}+U-\mu_{R}))]\bigg|{\frac{t_{L}t_{R}}{\varepsilon-E_{\alpha}}} - {\frac{t_{L}t_{R}}{\varepsilon - (E_{\alpha}+U)}} - {\frac{t_{L}t_{R}}{\varepsilon-E_{\beta}}} + {\frac{t_{L}t_{R}}{\varepsilon - (E_{\beta}+ U)}}\bigg|^2{d\varepsilon} \\
    \omega_{2,0}^{\Rightarrow} &= \frac{\pi\nu^{2}}{\hbar}{\int}[1-F_{FD}(\varepsilon-\mu_{R})]F_{FD}(\varepsilon-(E_{\alpha} + E_{\beta}+U-\mu_{R}))\bigg|{\frac{t_{R}t_{R}}{\varepsilon-E_{\alpha}}} - {\frac{t_{R}t_{R}}{\varepsilon - (E_{\alpha}+U)}} - {\frac{t_{R}t_{R}}{\varepsilon-E_{\beta}}} + {\frac{t_{R}t_{R}}{\varepsilon - (E_{\beta}+ U)}}\bigg|^2{d\varepsilon}, \\
    \omega_{2,0}^{\Leftarrow} &= \frac{\pi\nu^{2}}{\hbar}{\int}[1-F_{FD}(\varepsilon-\mu_{L})]F_{FD}(\varepsilon-(E_{\alpha} + E_{\beta}+U-\mu_{L}))\bigg|{\frac{t_{L}t_{L}}{\varepsilon-E_{\alpha}}} - {\frac{t_{L}t_{L}}{\varepsilon - (E_{\alpha}+U)}} - {\frac{t_{L}t_{L}}{\varepsilon-E_{\beta}}} + {\frac{t_{L}t_{L}}{\varepsilon - (E_{\beta}+ U)}}\bigg|^2{d\varepsilon},\\
    \omega_{2,0}^{\leftrightarrow} &= \frac{2\pi\nu^{2}}{\hbar}{\int}[1-F_{FD}(\varepsilon-\mu_{L})]F_{FD}(\varepsilon-(E_{\alpha} + E_{\beta}+U-\mu_{R}))\bigg|{\frac{t_{L}t_{R}}{\varepsilon-E_{\alpha}}} - {\frac{t_{L}t_{R}}{\varepsilon - (E_{\alpha}+U)}} - {\frac{t_{L}t_{R}}{\varepsilon-E_{\beta}}} + {\frac{t_{L}t_{R}}{\varepsilon - (E_{\beta}+ U)}}\bigg|^2{d\varepsilon}
\end{align}
\\
The rate expressions double charge-separated cotunneling processes are (my theory writing style changed):
\begin{align}
    \omega_{0,2}^{\Rightarrow} &= \frac{\pi\nu^{2}}{\hbar}{\int}F_{FD}(\varepsilon-\mu_{L})[1-F_{FD}(\varepsilon-(E_{\alpha}+E_{\beta}+U-\mu_{L}))]\bigg|{\frac{t_{L}t_{L}}{\varepsilon-E_{\alpha}}} - {\frac{t_{L}t_{L}}{\varepsilon - (E_{\alpha}+ U)}} - {\frac{t_{L}t_{L}}{\varepsilon-E_{\beta}}}+ {\frac{t_{L}t_{L}}{\varepsilon - (E_{\beta}+ U)}}\bigg|^2{d\varepsilon}, \\
    \omega_{0,2}^{\Leftarrow} &= \frac{\pi\nu^{2}}{\hbar}{\int}F_{FD}(\varepsilon-\mu_{R})[1-F_{FD}(\varepsilon-(E_{\alpha}+E_{\beta}+U-\mu_{R}))]\bigg|{\frac{t_{R}t_{R}}{\varepsilon-E_{\alpha}}} - {\frac{t_{R}t_{R}}{\varepsilon - (E_{\alpha}+U)}} - {\frac{t_{R}t_{R}}{\varepsilon-E_{\beta}}} + {\frac{t_{R}t_{R}}{\varepsilon - (E_{\beta}+ U)}}\bigg|^2{d\varepsilon},
\end{align}
\begin{align}
    \omega_{0,2}^{\leftrightarrow} &= \frac{2\pi\nu^{2}}{\hbar}{\int}F_{FD}(\varepsilon-\mu_{L})[1-F_{FD}(\varepsilon-(E_{\alpha}+E_{\beta}+U-\mu_{R}))]\bigg|{\frac{t_{L}t_{R}}{\varepsilon-E_{\alpha}}} - {\frac{t_{L}t_{R}}{\varepsilon - (E_{\alpha}+U)}} - {\frac{t_{L}t_{R}}{\varepsilon-E_{\beta}}} + {\frac{t_{L}t_{R}}{\varepsilon - (E_{\beta}+ U)}}\bigg|^2{d\varepsilon} \\
    &+ \frac{2\pi\nu^{2}}{\hbar}{\int}F_{FD}(\varepsilon-\mu_{R})[1-F_{FD}(\varepsilon-(E_{\alpha}+E_{\beta}+U-\mu_{L}))]\bigg|{\frac{t_{L}t_{R}}{\varepsilon-E_{\alpha}}} - {\frac{t_{L}t_{R}}{\varepsilon - (E_{\alpha}+U)}} - {\frac{t_{L}t_{R}}{\varepsilon-E_{\beta}}} + {\frac{t_{L}t_{R}}{\varepsilon - (E_{\beta}+ U)}}\bigg|^2{d\varepsilon}
\end{align}
\begin{align}
    \omega_{2,0}^{\Rightarrow} &= \frac{\pi\nu^{2}}{\hbar}{\int}[1-F_{FD}(\varepsilon-\mu_{R})]F_{FD}(\varepsilon-(E_{\alpha} + E_{\beta}+U-\mu_{R}))\bigg|{\frac{t_{R}t_{R}}{\varepsilon-E_{\alpha}}} - {\frac{t_{R}t_{R}}{\varepsilon - (E_{\alpha}+U)}} - {\frac{t_{R}t_{R}}{\varepsilon-E_{\beta}}} + {\frac{t_{R}t_{R}}{\varepsilon - (E_{\beta}+ U)}}\bigg|^2{d\varepsilon}, \\
    \omega_{2,0}^{\Leftarrow} &= \frac{\pi\nu^{2}}{\hbar}{\int}[1-F_{FD}(\varepsilon-\mu_{L})]F_{FD}(\varepsilon-(E_{\alpha} + E_{\beta}+U-\mu_{L}))\bigg|{\frac{t_{L}t_{L}}{\varepsilon-E_{\alpha}}} - {\frac{t_{L}t_{L}}{\varepsilon - (E_{\alpha}+U)}} - {\frac{t_{L}t_{L}}{\varepsilon-E_{\beta}}} + {\frac{t_{L}t_{L}}{\varepsilon - (E_{\beta}+ U)}}\bigg|^2{d\varepsilon},
\end{align}
\begin{align}
    \omega_{2,0}^{\leftrightarrow} &= \frac{2\pi\nu^{2}}{\hbar}{\int}[1-F_{FD}(\varepsilon-\mu_{L})]F_{FD}(\varepsilon-(E_{\alpha} + E_{\beta}+U-\mu_{R}))\bigg|{\frac{t_{L}t_{R}}{\varepsilon-E_{\alpha}}} - {\frac{t_{L}t_{R}}{\varepsilon - (E_{\alpha}+U)}} - {\frac{t_{L}t_{R}}{\varepsilon-E_{\beta}}} + {\frac{t_{L}t_{R}}{\varepsilon - (E_{\beta}+ U)}}\bigg|^2{d\varepsilon} \\
    &+ \frac{2\pi\nu^{2}}{\hbar}{\int}[1-F_{FD}(\varepsilon-\mu_{R})]F_{FD}(\varepsilon-(E_{\alpha} + E_{\beta}+U-\mu_{L}))\bigg|{\frac{t_{L}t_{R}}{\varepsilon-E_{\alpha}}} - {\frac{t_{L}t_{R}}{\varepsilon - (E_{\alpha}+U)}} - {\frac{t_{L}t_{R}}{\varepsilon-E_{\beta}}} + {\frac{t_{L}t_{R}}{\varepsilon - (E_{\beta}+ U)}}\bigg|^2{d\varepsilon}
\end{align}
\end{widetext}

\bibliography{References}
\end{document}